\documentclass[11pt]{article}
\pdfoutput=1
\usepackage[T1]{fontenc}
\usepackage[latin1]{inputenc}
\usepackage[nosort]{cite}
\usepackage{color}
\usepackage[bulletsep]{collref}
\usepackage{graphicx}
\usepackage{bbm}
\usepackage{amsmath}
\usepackage{amssymb}
\usepackage{subfig}
\usepackage{multirow}
\usepackage{tikz}
\usepackage{hyperref}

\makeatletter \@addtoreset{equation}{section} \makeatother

\makeatletter
\let\old@startsection=\@startsection
\let\oldl@section=\l@section
\renewcommand{\@startsection}[6]{\old@startsection{#1}{#2}{#3}{#4}{#5}{#6\mathversion{bold}}}
\renewcommand{\l@section}[2]{\oldl@section{\mathversion{bold}#1}{#2}}
\makeatother

\makeatletter
\let\old@makecaption=\@makecaption
\def\@makecaption{\small\old@makecaption}

\makeatother

\let\oldPhi=\Phi
\let\oldPsi=\Psi
\let\oldGamma=\Gamma
\let\oldDelta=\Delta
\let\oldSigma=\Sigma
\let\oldTheta=\Theta
\let\oldPi=\Pi
\let\oldUpsilon=\Upsilon
\renewcommand{\Phi}{\mathnormal{\oldPhi}}
\renewcommand{\Psi}{\mathnormal{\oldPsi}}
\renewcommand{\Gamma}{\mathnormal{\oldGamma}}
\renewcommand{\Sigma}{\mathnormal{\oldSigma}}
\renewcommand{\Delta}{\mathnormal{\oldDelta}}
\renewcommand{\Theta}{\mathnormal{\oldTheta}}
\renewcommand{\Pi}{\mathnormal{\oldPi}}
\renewcommand{\Upsilon}{\mathnormal{\oldUpsilon}}

\renewcommand{\Im}{\mathop{\mathrm{Im}}}

\newcommand{\Sphere}{\mathrm{S}}  
\newcommand{\AdS}{\mathrm{AdS}}

\ifx\genfrac\sdflkaj

\else

\fi

\newcommand{\p}{\partial}

\def\[{\begin{equation}}
\def\]{\end{equation}}

\makeatletter
\def\mr@ignsp#1 {\ifx\:#1\@empty\else #1\expandafter\mr@ignsp\fi}%
\newcommand{\multiref}[1]{\begingroup
\xdef\mr@no@sparg{\expandafter\mr@ignsp#1 \: }%
\def\mr@comma{}%
\@for\mr@refs:=\mr@no@sparg\do{\mr@comma\def\mr@comma{,}\ref{\mr@refs}}%
\endgroup}
\makeatother

\newcommand{\hypref}[2]{\ifx\href\asklfhas #2\else\href{#1}{#2}\fi}

\renewcommand{\eqref}[1]{(\multiref{#1})}

\ifx\href\asklfhas\newcommand{\href}[2]{#2}\fi

\newcommand{\be}{\begin{eqnarray}}
\newcommand{\ee}{\end{eqnarray}}

\newcommand{\Elliptic}[1]{\mathbbm{#1}}
\newcommand{\EllipticE}{\Elliptic{E}}

\newcommand{\EllipticK}{\Elliptic{K}}

\DeclareMathOperator{\JacobiAM}{am}

\DeclareMathOperator{\JacobiCN}{cn}
\DeclareMathOperator{\JacobiDN}{dn}
\DeclareMathOperator{\JacobiSN}{sn}

\DeclareMathOperator{\JacobiDS}{ds}

\begin{document}

\thispagestyle{empty}
\begin{flushright}\footnotesize
\texttt{NORDITA-2016-109}\\
\end{flushright}
\vspace{1cm}

\begin{center}%
{\Large\textbf{\mathversion{bold}%
Dual conformal transformations \\
of smooth holographic Wilson loops
}\par}

\vspace{1.5cm}

\textrm{Amit Dekel} \vspace{8mm} \\
\textit{%
Nordita, KTH Royal Institute of Technology and Stockholm University, \\
Roslagstullsbacken 23, SE-106 91 Stockholm, Sweden
} \\

\texttt{\\ amit.dekel@nordita.org}

\par\vspace{14mm}


\begin{abstract}\noindent
We study dual conformal transformations of minimal area surfaces in $\AdS_5\times \Sphere^5$ corresponding to holographic smooth Wilson loops and some other related observables.
To act with dual conformal transformations we map the string solutions to the dual space by means of T-duality, then we apply a conformal transformation and finally T-dualize back to the original space.
The transformation maps between string solutions with different boundary contours.
The boundary contours of the minimal surfaces are not mapped back to the AdS boundary, and the regularized area of the surface changes.
\end{abstract}

\begin{minipage}{14cm}

\end{minipage}

\end{center}

\newpage

\tableofcontents

\bigskip
\noindent\hrulefill
\bigskip


\section{Introduction}

Wilson loops are fundamental observables in any gauge theory.
In gauge theories such as $\mathcal{N}=4$ SYM, we are able to explore these objects at strong coupling using holography, by studying minimal surfaces in the dual geometry \cite{Maldacena:1998im,Rey:1998ik}.
Finding these minimal surfaces and calculating their area, which corresponds to the expectation value of the Wilson loops, is a challenging mathematical problem.

The problem in AdS space is of particular interest since the $\sigma$-model is classically integrable.
Thus, it should be possible to find exact results and perhaps solve the general problem by the advanced tools of integrability.
During the past several years various methods were proposed and applied for calculating minimal surfaces in AdS space and their area.
To date, a general family of solutions in $\AdS_3$ is known in terms of Riemann theta-functions, where one starts with an algebraic curve and generates a solutions \cite{Babich:1992mc,Ishizeki:2011bf}.
Another approach is to reformulate the problem in terms of finding the boundary curve parametrization in a given gauge \cite{Kruczenski:2014bla}, which allowed to make some progress \cite{Dekel:2015bla,Huang:2016atz}.
In a different approach one takes the smooth continuous limit of lightlike Wilson loops \cite{Toledo:2014koa}, where the solution to the problem is known in terms of a TBA-like integral equation \cite{Alday:2010vh}.
Other known integrability techniques were used to generate new solutions from known solutions \cite{Kalousios:2011hc}.

In parallel there is a growing interest in the symmetry group of Wilson loops in $\mathcal{N}=4$ SYM, which is expected be larger than the obvious conformal symmetry group, where there is evidence that the symmetry algebra extends to Yangian symmetry algebra \cite{Muller:2013rta,Munkler:2015gja,Munkler:2015xqa,Beisert:2015uda}.
In the case of lightlike Wilson loops the extended symmetry, also known as the dual conformal symmetry, is related to scattering amplitudes which are the dual objects under a sequence of bosonic and fermionic T-dualities, under which the $\AdS_5\times \Sphere^5$ background is self-dual \cite{Berkovits:2008ic,Beisert:2008iq}.
The dual conformal generators are related to the usual conformal generators in the T-dual space \cite{Alday:2007hr,Berkovits:2008ic,Beisert:2008iq,Beisert:2009cs}.
It is thus interesting to study how dual conformal transformations act on smooth Wilson loops, which is what we shall do in this paper.

The procedure we use is to first T-dualize a classical solution along the flat directions of the AdS space in the Poincare patch, then act with a conformal transformation in the dual space and finally T-dualize back to the original space.
The procedure is guaranteed to yield a solution to the equations of motion, which in principle can be different from the original one (namely, not related by conformal transformations).
We shall call this action a symmetry if it leaves the expectation value (or equivalently regularized area) invariant.
The procedure can be applied for any string solution (not necessarily a solution corresponding to a Wilson loop) such as correlation functions of Wilson loops, operators etc. which are also described by minimal surfaces in $\AdS_5\times \Sphere^5$ space.

We are going to show how the boundary contour and minimal area surface change under these transformations and how the expectation value changes in some of the cases.
We shall see that the transformation generates new solutions with different boundary contour, area and sometimes even topology compared to the original solutions.
Moreover, the surface has to be analytically continued since the boundary contour will not be mapped to the AdS boundary.

The dual conformal transformation is non-local and requires the knowledge of the minimal surface solution and not just boundary contour.
Since the general minimal area surface solution for any boundary contour is not known, we are restricted to study the transformation of known solutions.
A natural starting point is the 1/2 BPS straight Wilson line (or circular Wilson loop) which is the simplest solution in $\AdS_3$, which we analyze first. Then, we continue to the wavy line (or circle) in $\AdS_3$ where things get more interesting.
More general solutions in $\AdS_3$ are typically of higher genus (in terms of the algebraic curve and the Riemann theta-functions) and so are more complicated in nature. We study the genus-1 solution corresponding to the $q \bar q$-potential (for example, other interesting genus-1 solutions are the cusp and the correlation function of two Wilson loops). We do not analyze higher genus solutions in this paper.
Afterwards, we continue to study some string solutions in higher dimensions.
We study the longitude solution which lives in $\AdS_4\times \Sphere^2$ \cite{Drukker:2007qr}, the correlation function between a circular Wilson loop and a BMN operator $\left<W \mathcal{O}_\mathcal{J}\right>$ solution which lives in $\AdS_3\times \Sphere^1$ \cite{Zarembo:2002ph} and the BMN geodesic \cite{Dobashi:2002ar,Tsuji:2006zn,Janik:2010gc}.
We begin by explaining the general transformation procedure, then apply it to the various solutions mentioned above, and we end with a discussion.

\section{Dual conformal transformations}

In this section we are going to describe the procedure of acting with dual conformal transformations on classical string solutions in $\AdS_5 \times \Sphere^5$.
The transformation that leaves the $\AdS_5 \times \Sphere^5$ background invariant involves a sequence of bosonic and fermionic T-dualities, however since our discussion is classical, we shall only be concerned with the bosonic part\footnote{We ignore the fermionic T-dualities which compensating for the shift of the dilaton, but do not change the metric. For similar reasons $\AdS_n \times \Sphere^n$ $(n=2,3,5)$ were shown to be self-dual, whereas the case of $\AdS_4 \times \mathbb{C}P^3$ so far has not \cite{Berkovits:2008ic,Adam:2009kt,Adam:2010hh,Dekel:2011qw}. See also \cite{Abbott:2015ava,Abbott:2015mla,Colgain:2016gdj} for more recent developments.}.
Thus, we may consider solutions in AdS space in any dimension, and concentrate on the AdS part only.

We shall work in the Poincare patch $ds^2 = \frac{dX^\mu dX_\mu  +dZ^2}{Z^2}$, and start by T-dualizing along all the flat directions $X^\mu$.
The T-duality is followed by a field redefinition $Z\to 1/Z$ in order to get back an AdS background in the same form.
Next we apply a conformal transformation on the dual solution and then T-dualize back to the "original" space.
Throughout the paper we are going to denote the original solution by $X^\mu,Z$ and the transformed one using hatted coordinates $\hat{X}^\mu,\hat{Z}$.
In the following we explain how the procedure works in more detail.

\subsection*{T-duality}
Let us start with a classical string solution given in Poincare coordinates $X^\mu$ and $Z$, and assume the worldsheet is Euclidean.
The solution solves the equations of motion
\begin{align}
d\left(\frac{1}{Z^2}\ast d X^\mu\right) = 0, \quad
d\ast d \ln Z + \frac{1}{Z^2}d X^\mu\wedge \ast dX_\mu = 0,
\end{align}
where '$\ast $' is the Hodge dual operator on the worldsheet.
T-dualizing along the $X^\mu$ directions gives a set of new coordinates given by
\begin{align}\label{eq:Tduality}
d\tilde X^\mu = \frac{i}{Z^2}\ast d X^\mu, \quad
\tilde Z = \frac{1}{Z},
\end{align}
where the factor of $i$ is present due to our assumption of Euclidean worldsheet (otherwise, for Minkowsian worldsheet the factor is absent).
The new fields satisfy the same equations of motion as the original fields, and for the Euclidean case we see that T-duality will map real solutions to complex solutions.
More explicitly, in components we have
\begin{align}
\p_\tau \tilde X^\mu = \frac{i}{Z^2}\p_\sigma X^\mu, \quad
\p_\sigma \tilde X^\mu = - \frac{i}{Z^2}\p_\tau X^\mu, \quad
\tilde Z = \frac{1}{Z}.
\end{align}

\subsection*{Dual conformal transformation}
The next step is to apply conformal transformations to the solution in the dual space.
It is easy to see that translations, rotations and dilatations of the dual solution correspond to conformal transformations of the original solution.
However, special conformal transformation which act by
\begin{align}
\tilde X^\mu \to \tilde X^{' \mu } = \frac{\tilde X^\mu + b^\mu (\tilde X^2+\tilde Z^2)}{1+2 b\cdot \tilde X + b^2 (\tilde X^2 + \tilde Z^2)},\quad
\tilde Z \to \tilde Z^{'} = \frac{\tilde Z}{1+2 b\cdot \tilde X + b^2 (\tilde X^2 + \tilde Z^2)},
\end{align}
generally do not correspond to ordinary conformal transformations of the original solution.
For the Euclidean case, in order to end up with a real solution after T-dualizing back, we need the parameter $b^\mu$ to be purely imaginary, so we will take $b^\mu \to i b^\mu$.

\subsection*{T-dualizing back}

The final step is to T-dualize back to the original space using (\ref{eq:Tduality}), where $X^\mu ,Z$ are replaced by $\tilde X'^\mu ,\tilde Z'$, and $\tilde X^\mu ,\tilde Z$ by $\hat{X}^\mu ,\hat{Z}$.
The resulting solution is a second order polynomial in $b=|b^\mu|$, namely
\begin{align}
\hat{X}^\mu \to X_\mu + b^\nu Y_\nu^\mu + b^2 \chi^\mu,\quad
\hat{Z} \to Z + b^\nu Y_\nu + b^2 \zeta,
\end{align}
where $\{X_\mu,Z\}$ is the original solution, and the structure implies that the set $\{\chi_\mu,\zeta\}$ is also a solution to the equations of motion.
Let us stress that $\chi_\mu$ and $\zeta$ depend on the unit vector $\hat b^{\mu}$.
The reason we get this structure for $\hat Z$ is obvious, and for $\hat X^\mu$ one can check explicitly that the potential third order in $|b|$ contribution vanishes.

\section{Analysis}

In this section we act with dual conformal transformations on string solutions according to the procedure described in the previous section.
We start by studying some solutions in Euclidean $\AdS_3$ which include the 1/2 BPS Wilson loops, the wavy line and the $q \bar q$-potential.
Then we continue to string configurations living in higher dimensions.
These include the longitude solution, the $\left<W \mathcal{O}_{\mathcal{J}}\right>$ solution and the BMN geodesic.

\subsection{The straight Wilson line and circular Wilson loop}

\subsubsection*{The straight line}
The simplest Wilson loop available in $\AdS_3$ is the infinite straight Wilson line.
The string solution is given by
\begin{align}
X_1 = \sigma,\quad
X_{\mu \neq 1} = 0,\quad
Z = \tau,
\end{align}
where $\tau>0$ and $-\infty<\sigma<\infty$.
The dual conformal transformation with the parameter $b^\mu = (b_1,b_2)$ maps the solution to
\begin{align}
\hat{X}_1 = \sigma,\quad
\hat{X}_{\mu \neq 1} = 0,\quad
\hat{Z} = \tau + 2 b_1.
\end{align}

Thus, we got back the same solution with a different range of the $\tau$ coordinate, which is now $-2 b_1 <\tau<\infty$.
This is quite trivial, but we should notice that when $b_1>0$ we need to continue the surface
to include the part that was mapped to $Z<0$ before the transformation.
Now we need to define a new regulator at $\hat{Z} = \epsilon$, so on the worldsheet it is defined at
$\tau = \epsilon - 2 b_1$.

Let us stress that the Lagrangian density has changed,
so if we were to keep track of the original regulator at $\tau = \epsilon$ we would end with a different area.

\subsubsection*{The circle}

\begin{figure}
    \centering
    \includegraphics[trim = 0mm 0mm 0mm 0mm,clip,width = 0.3\textwidth]{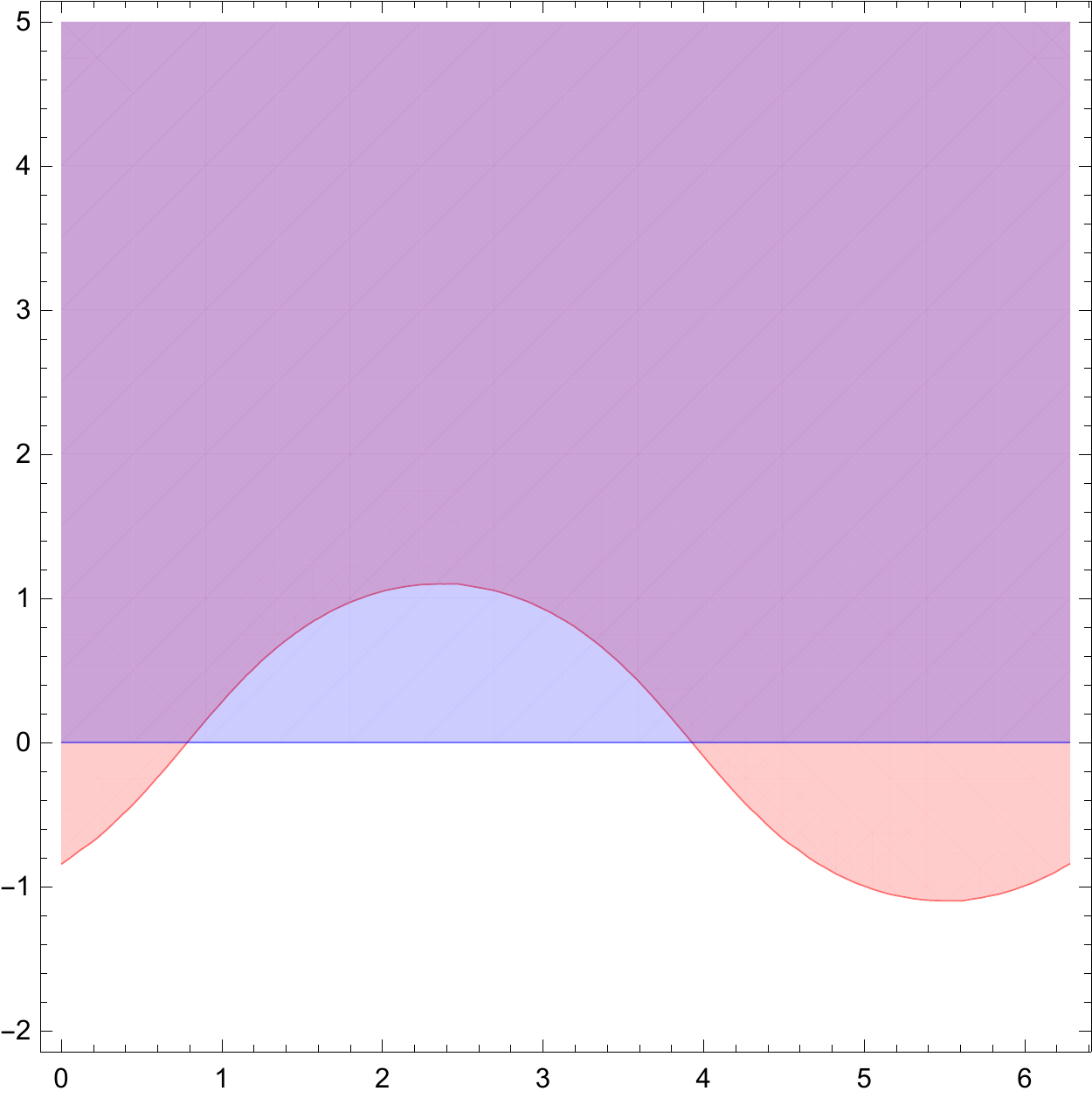}
    \caption{In this plot we show the part of the worldsheet which is mapped to the circular Wilson loops solution.
    Before the transformation the minimal surface is mapped from the semi-infinite strip which we mark in blue, and the boundary contour is the $\tau = 0$ line. After the dual conformal transformation the minimal surface is mapped from the light red area. The two section partialy overlap (denoted in purple) and the boundaries meet at the AdS boundary at two points.}
    \label{fig:WSmapToCircb05betapio4}
\end{figure}

The same analysis can be carried for the circular solution
\begin{align}
X_1 = \frac{\cos\sigma}{\cosh\tau},\quad
X_2 = \frac{\sin\sigma}{\cosh\tau},\quad
Z = \tanh\tau,
\end{align}
where $0\leq\tau<\infty$ and $0\leq\sigma<2\pi$.
After the transformation we get
\begin{align}
\hat{X}_1 = ~&\frac{\cos\sigma}{\cosh\tau} -2 b \sin \beta \tanh\tau + b^2 \text{sech}\tau \cos (2 \beta -\sigma ) ,\nonumber\\
\hat{X}_2 = ~&\frac{\sin\sigma}{\cosh\tau} +2 b \cos \beta \tanh \tau + b^2 \text{sech}\tau \sin (2 \beta -\sigma ),\nonumber\\
\hat{Z} = ~&\tanh \tau+2 b \frac{\sin (\beta -\sigma )}{\cosh\tau} -b^2\tanh \tau,
\end{align}
where $b^\mu = b(\cos\beta,\sin\beta)$.
Again we get the same surface in a different parametrization.
As for the line, the original boundary contour defined for $\tau = 0$ is not mapped to the AdS boundary, but rather intersects the AdS boundary at two points, $\tau=0$ and $\sigma = \beta,\pi + \beta$.
Also the new boundary contour is mapped from a different part of the worldsheet, see figure \ref{fig:WSmapToCircb05betapio4}.

The infinite line and circular solutions are related by a conformal transformation.
We see that in both case the "complete" surface (namely, including $Z\geq0$ and $Z<0$) transforms into itself, however the original boundary curves transform differently. In the circular case it has two parts, mapped to positive and negative $Z$,
while in the case of the line the whole boundary is mapped either positive or negative value of $Z$. This difference is due to the fact that dual conformal transformations do not commute with ordinary conformal transformations.

\subsection{The wavy line}
Our next step is to perturb the previous result by a adding a small localized deformation to the straight line.
The minimal surface as well as the regularized area are known in terms of an integral of the deformation function $\xi(\sigma)$ \cite{Polyakov:2000ti,Semenoff:2004qr}.
Concretely we have
\begin{align}
X_1 = \sigma,\quad
X_2 = \epsilon \xi(\tau,\sigma),\quad
Z = \tau,
\end{align}
where $\epsilon$ is a small parameter such that $|\epsilon \xi(\sigma,\tau)|<<1$.
The function $\xi(\sigma,\tau)$ is given by
\begin{align}
\xi(\sigma,\tau) = \frac{2\tau^3}{\pi}\int_{-\infty}^{\infty}d\sigma'\frac{\xi(\sigma')}{((\sigma-\sigma')^2 + \tau^2)^2}.
\end{align}
The dual conformal transformation maps the infinite line to itself, and is expected to map the wavy line to another wavy line, defined by some other function $\xi'(\sigma)$.
The wavy line as expressed above is defined for $\tau>0$ and $-\infty<\sigma<\infty$.
By the integral definition of $\xi(\sigma,\tau)$ we see that the surface is not continued smoothly for $\tau<0$ since the pole at $\sigma' = \sigma + i \tau$ moves below the real axis.
In order to proceed with our prescription it is essential that we have a smooth surface when $\tau$ becomes negative.
A natural way to proceed is to consider the resulting function $\xi(\sigma,\tau)$ for $\tau>0$ also for $\tau\leq 0$.

Acting with our non-local transformation yields
\begin{align}
&\hat{X}_1 = \sigma + \epsilon  \frac{2 b \sin \beta }{\pi }\int \frac{ \left((\sigma -\sigma')^2 - \tau ^2\right)}{\left((\sigma -\sigma')^2 + \tau ^2\right)^2} \xi (\sigma')\, d\sigma',\nonumber\\
&\hat{X}_2 = \epsilon \xi(\sigma,\tau) -  \epsilon\frac{2 b \cos\beta}{\pi }\int \frac{\left((\sigma -\sigma')^2 - \tau ^2\right)}{\left((\sigma -\sigma')^2 + \tau ^2\right)^2} \xi (\sigma')\, d\sigma',\nonumber\\
&\hat{Z} = \tau  + 2 b \cos \beta
-\epsilon  \frac{4 \tau b \sin \beta }{\pi }\int \frac{ \left(\sigma -\sigma'\right)}{\left((\sigma -\sigma')^2 + \tau ^2\right)^2} \xi (\sigma')\, d\sigma'.
\end{align}
Notice that $b_1$ is assumed to be negative since we assumed that $\tau>0$ and so in order for $\hat{Z}\geq0$ we will assume $b_1<0$.
In this way, to leading order $-2 b_1<\tau<\infty$, so the new deformation is still small for large $|b|$ since it implies that $\tau$ is also large.

Next we restrict ourselves to the new boundary curve ($\hat{Z}=0$ curve), which we can reparameterize to have the form $(\sigma,\epsilon \hat{\xi}(\sigma))$.
The result is
\begin{align}
\hat{\xi}(\sigma) \equiv T_{\vec b}[\xi(\sigma)] =  - \frac{2 b_1}{\pi}\int d\sigma' \frac{\xi(\sigma')}{\left((\sigma -\sigma')^2 + (2 b_1)^2\right)}.
\end{align}
An immediate consequence of the above equation is that the resulting curve depends only on one parameter, $b_1 = b \cos\beta$ and not on $b_2 = b \sin\beta$.
As a check we can also consider now two consecutive transformations, which result in
\begin{align}
T_{\vec b}[T_{\vec a}[\xi(\sigma)]] = T_{\vec b + \vec a}[\xi(\sigma)],
\end{align}
so as one expects, two such transformations amount to the same transformation with different parameters.
For the wavy line, the contribution to the regularized area is
\begin{align}
A_{\text{reg}} = -\frac{\epsilon^2}{4\pi}\int d\sigma\int d\sigma' \frac{(\partial_\sigma\xi(\sigma)-\partial_{\sigma'}\xi(\sigma'))^2}{(\sigma-\sigma')^2}.
\end{align}
The resulting deformation $\xi'$ changes the value of this area functional.
In order to get a finite contribution to the area the deformation should be localized (possibly at several points) and fall off at infinity.

The conformally invariant information is encoded in the holomorphic function $f(z)$ which appears when we Pohlmeyer reduce the model to the generalized sinh-Gordon model (here $z=\sigma + i \tau$) \cite{Kruczenski:2014bla}. In the conformal gauge, when we map the solution from the upper half plane on the worldsheet, the relation is given by \cite{Kruczenski:2014bla,Dekel:2015bla}
\begin{align}
\partial^3_\sigma\xi(\sigma) = -4 \Im f(\sigma).
\end{align}

Let us consider explicit simple example where $\xi(\sigma) = \frac{1}{1+\sigma^2}$.
The resulting deformation after the transformation is given by
\begin{align}
\hat{\xi}(\sigma) = \frac{(1-2 b_1)}{(1-2 b_1)^2 + \sigma^2}.
\end{align}
The regularized area of the new curve is given by
\begin{align}
A_{\text{reg}} = -\epsilon^2\frac{3 \pi }{16(1-2 b_1)^4},
\end{align}
which clearly depends on $b$. Since $b_1<0$ the area grows, and in the limit $b_1 \to -\infty$ it goes to zero, where we recover the straight line.
This is a general feature as can be seen from the general expression, though this limit is not well defined since also the lower boundary of of the $\tau$ interval will go to infinity.

In this example the holomorphic function before and after the transformation is given by $f(z) = -\frac{3\epsilon}{2(z+i)^4}$ and $f(z) = -\frac{3\epsilon}{2(z+i(1-2 b_1))^4}$ respectively\footnote{For the case of mapping the minimal surface from the unit disk on the worldsheet as in \cite{Kruczenski:2014bla}, these $f(z)$ functions are given by $-\frac{3\epsilon}{8}$ and $-\frac{3\epsilon}{8(i+b_1(z-i))^4}$ respectively.}.

\subsection{Deformation of the circle beyond the wavy approximation}

Using the method introduced in \cite{Dekel:2015bla} we can explore deformations of the line or the circle beyond the leading order.
In contrast to the case of the wavy approximation we do not know the general solution for an arbitrary perturbation, but have to solve for specific examples.
One way is to deform the contour directly, but different deformations may be related by conformal transformations.
Here we prefer to deform the holomorphic function $f(z)$ which vanished for the circle, and encodes the conformally invariant data once the boundary on the worldsheet is fixed (the unit circle in our case), see \cite{Kruczenski:2014bla,Dekel:2015bla} for more details.

One of the simplest ways to perturb the circle is by using $f(z) = \epsilon e^{i \varphi}$, namely a complex constant\footnote{Actually the exact regularized area is known for any $f(z) = f_0 z^n$ with $n\geq 0$ an integer \cite{Huang:2016atz}. The constant phase $\varphi$ is related to the spectral parameter and generally defines a family of solutions which are not related by conformal symmetry that have the same area \cite{Kruczenski:2014bla}. In this case $\varphi$ acts simply by rotation of the boundary curve, namely it is a conformal transformation.}, where the exact solution for the regularized area is known \cite{Huang:2016atz}, though the minimal surface is not.
It is easy to show that for these boundary curves the spectral parameter $\varphi$-deformation acts by rotations, and does not change the area (namely it corresponds to a conformal transformation in this simple case).
In this section we carry the analysis for this Wilson loop to 4th order in $\epsilon$.

\begin{figure}
    \centering
  \includegraphics[width=60mm]{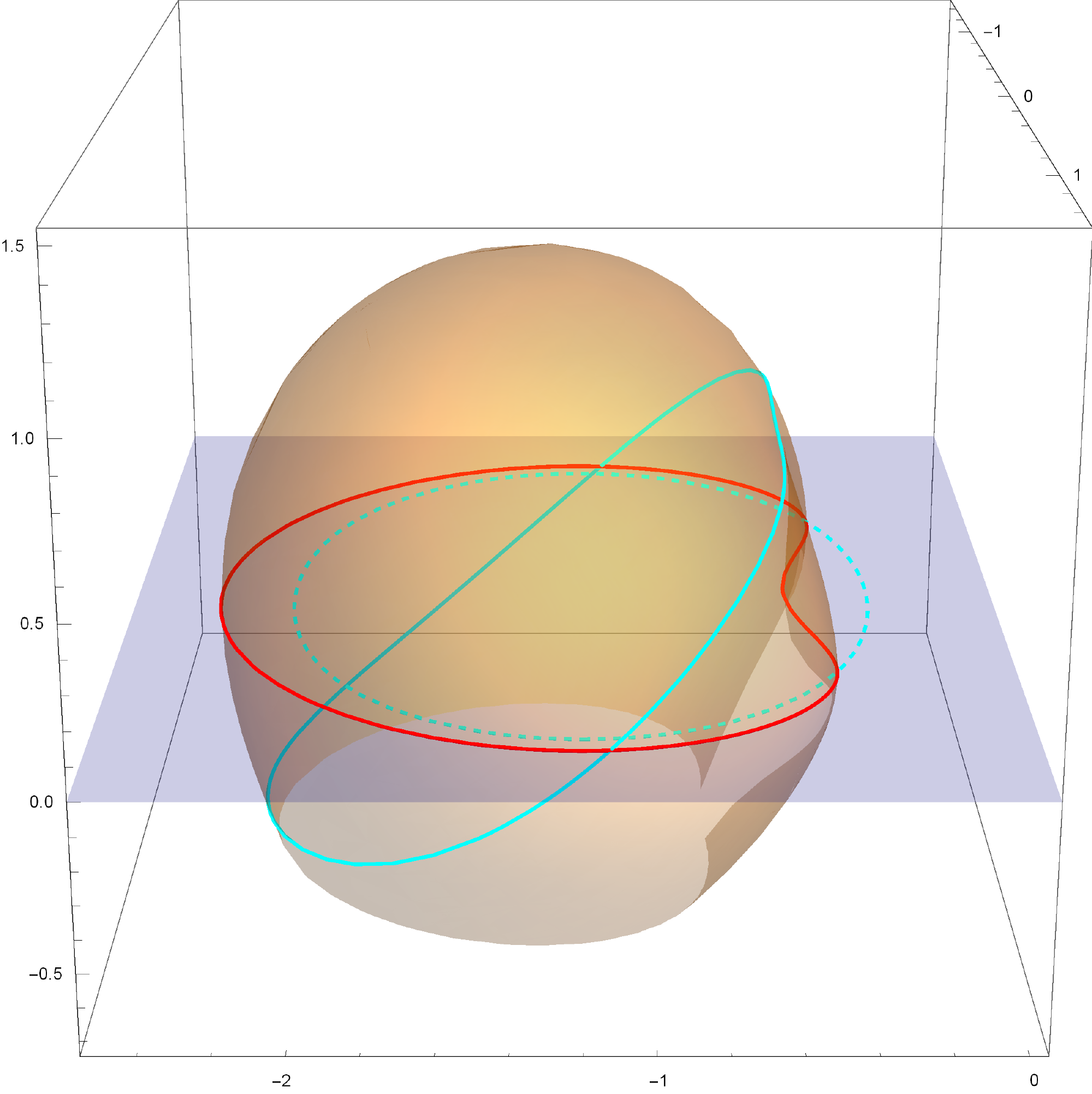} \hspace{3mm}
    \caption{The $f(z) = \epsilon e^{i \varphi}$ solution for $\epsilon=0.2$, $b=0.3$, $\beta=0$ $\phi = 0$ in $\AdS_3$.
    The dashed curve corresponds to the original boundary curve (up to translations), the red contour shows the new boundary curve and the cyan contour corresponds to the transformed original contour. The new Wilson loop minimal surface corresponds to the surface above the blue surface which is defined by the AdS boundary.}
    \label{fig:fconstexample}
\end{figure}

The first task is to find the minimal surface solution analytically, which can be done by first solving the generalized cosh-Gordon equation to 4th order, then solving the axillary linear problem and then extract the solution for $(X_1,X_2,Z)$.
Then we can apply our procedure, the solution is quite long and messy and we do not present it here explicitly.
After applying the dual conformal transformation the original boundary curve is not mapped to the boundary, however, we can identify a new closed contour on the worldsheet which is mapped to the boundary, see figure \ref{fig:fconstexample}. Throughout the procedure the function $f(z)$ does not change, but since the boundary contour on the worldsheet is changed, the function $\alpha(z,\bar z)$ changes. Remember that $\alpha$ has to diverge at the boundary contour, which happens on the unit circle for our initial solution.

In order to study the new solution it is convenient to map the worldsheet region which is mapped to the minimal surface, back to the unit disk using a holomorphic map, $z=h(w)$. Then our new $f(w)$ follows by $f_{\text{new}}(w) = (\p_w h(w))^2 f(h(w))$, and we can easily solve for $\alpha_{\text{new}}(w,\bar w)$, and compute the regularized area.
\begin{align}
A_{ren}^{\text{new}} = -2\pi - 4 \int_0^{2\pi} d\theta \int_0^1 dr r |f_{\text{new}}|^2 e^{-2 \alpha_{\text{new}}}.
\end{align}

Before the transformation the area is given by
\begin{align}
A_{ren} = -2\pi -\frac{4 \pi  \epsilon^2}{3}-\frac{44 \pi  \epsilon^4}{135}-\frac{1504 \pi  \epsilon^6}{8505}+O\left(\epsilon^7\right),
\end{align}
while after we have
\begin{align}
A_{ren}^{\text{new}} = &-2\pi -\frac{4 \pi  \left(b^2+1\right)^4 \epsilon^2}{3 \left(b^2-1\right)^4}
-\frac{128 \pi  b^2 \left(b^2+1\right)^3 \left(9 b^4+2 b^2-3\right) \cos (2 \beta -\varphi )\epsilon^3}{9 \left(b^2-1\right)^7}\nonumber\\
&+\frac{4 \pi  \left(b^2+1\right)^2 \epsilon^4}{135 \left(b^2-1\right)^{10}}
\bigg(-192 \left(845 b^8+532 b^6-562 b^4-204 b^2+125\right) b^4 \cos (4 \beta -2 \varphi )\nonumber\\
&-1163 b^{16}-41260 b^{14}-210668 b^{12}-140756 b^{10}+42350 b^8+17452 b^6-10220 b^4\nonumber\\
&+212 b^2-11\bigg)
+O\left(\epsilon^5\right).
\end{align}
Clearly the area (and so the expectation value) has changed already at the lowest order of the perturbation.
This immediately implies that the contours are not related by conformal symmetry.
Starting at the third order we see $\varphi$-dependence which reflects the fact that the transformation does not commute with some of the original conformal transformations (where in this case $\varphi$ corresponds to rotations of the original solution, and more generally $\varphi = \pi/2$ is always a conformal transformation).
Moreover, to leading order the original curve is related to our previous example $\xi(\sigma) = \frac{1}{1+\sigma^2}$ by a conformal transformation. However, already the $\epsilon^2$ coefficient has different $b$ dependence which again is related to the fact that the transformation does not commute with ordinary conformal transformations.

In this expansion  we should keep $\epsilon$ small, but so does $b$ has to be small where otherwise we will be taken away from the wavy regime as can be seen for example by the form of the leading correction for the renormalized area.
To leading order $\epsilon$ is just rescaled by the factor $\frac{\left(b^2+1\right)^2}{\left(b^2-1\right)^2}$.
This argument implies that we can also use large $b$, however while the higher terms are finite for $b\to \infty$ they (the even ones) grow very fast so $\epsilon$ should be even smaller.

\subsection{$q\bar q$ potential}
The last solution we study in Euclidean $\AdS_3$ is the $q\bar q$ potential solution solution given by \cite{Chu:2009qt,Janik:2012ws}
\begin{align}\label{eq:qqbarsol}
X_1 = \sigma,\quad
X_2 = \tau - \EllipticE(\JacobiAM(\tau,-1),-1),\quad
Z   = \JacobiSN(\tau,-1),
\end{align}
where '$\JacobiSN$' is a Jacobi elliptic function, '$\EllipticE$' is the elliptic integral of the second kind and '$\JacobiAM$' the Jacobi amplitude.
This is a genus-1 solution (in terms of the algebraic curve) which is reflected by the presence of the elliptic functions above (equivalently genus-1 Riemann theta functions).
After the transformation we get
\begin{align}
\hat{X}_1 = ~&
\sigma
-2 b_1 \sigma \tilde\EllipticE
+b_2 \left(-\sigma^2+\tilde\EllipticE^2+\JacobiSN^2-1\right)\nonumber\\
&+\frac{1}{3} \left(b_2^2-b_1^2\right) \sigma \left(\sigma^2-3 \left(\tilde\EllipticE^2+\JacobiSN^2\right)\right)
+\frac{2}{3} b_1 b_2 \left(-2 \tau -3 \tilde\EllipticE (\JacobiSN^2 - \sigma^2  ) - \tilde\EllipticE^3+2 \JacobiSN \JacobiCN\JacobiDN\right)
\nonumber\\
\hat{X}_2 = ~&
\tilde\EllipticE
-2 b_2 \sigma \tilde\EllipticE
+b_1 \left(\sigma^2-\tilde\EllipticE^2-\JacobiSN^2+1\right)\nonumber\\
&
+\frac{1}{3} \left(b_2^2-b_1^2\right) \left(-2 \tau -3 \tilde\EllipticE (\JacobiSN^2 -\sigma^2 )-\tilde\EllipticE^3+2 \JacobiSN \JacobiCN\JacobiDN\right)
+\frac{2}{3}b_1 b_2 \left(3 \sigma \left(\tilde\EllipticE^2+\JacobiSN^2\right)-\sigma^3\right)
\nonumber\\
\hat{Z}   = ~&
\JacobiSN+2 b_1 \left(\JacobiCN\JacobiDN-\tilde\EllipticE \JacobiSN\right)-2 b_2 \sigma \JacobiSN
+\left(b_1^2+b_2^2\right) \left(-2 \tilde\EllipticE \JacobiCN\JacobiDN+\sigma^2 \JacobiSN+\tilde\EllipticE^2 \JacobiSN-\JacobiSN^3\right)
,
\end{align}
where we used the short hand notation $\tilde\EllipticE \equiv \tau - \EllipticE(\JacobiAM(\tau,-1),-1)$, and $\JacobiSN \equiv \JacobiSN(\tau,-1)$ etc.

\begin{figure}
    \centering
    \subfloat[][]{
  \includegraphics[width=50mm]{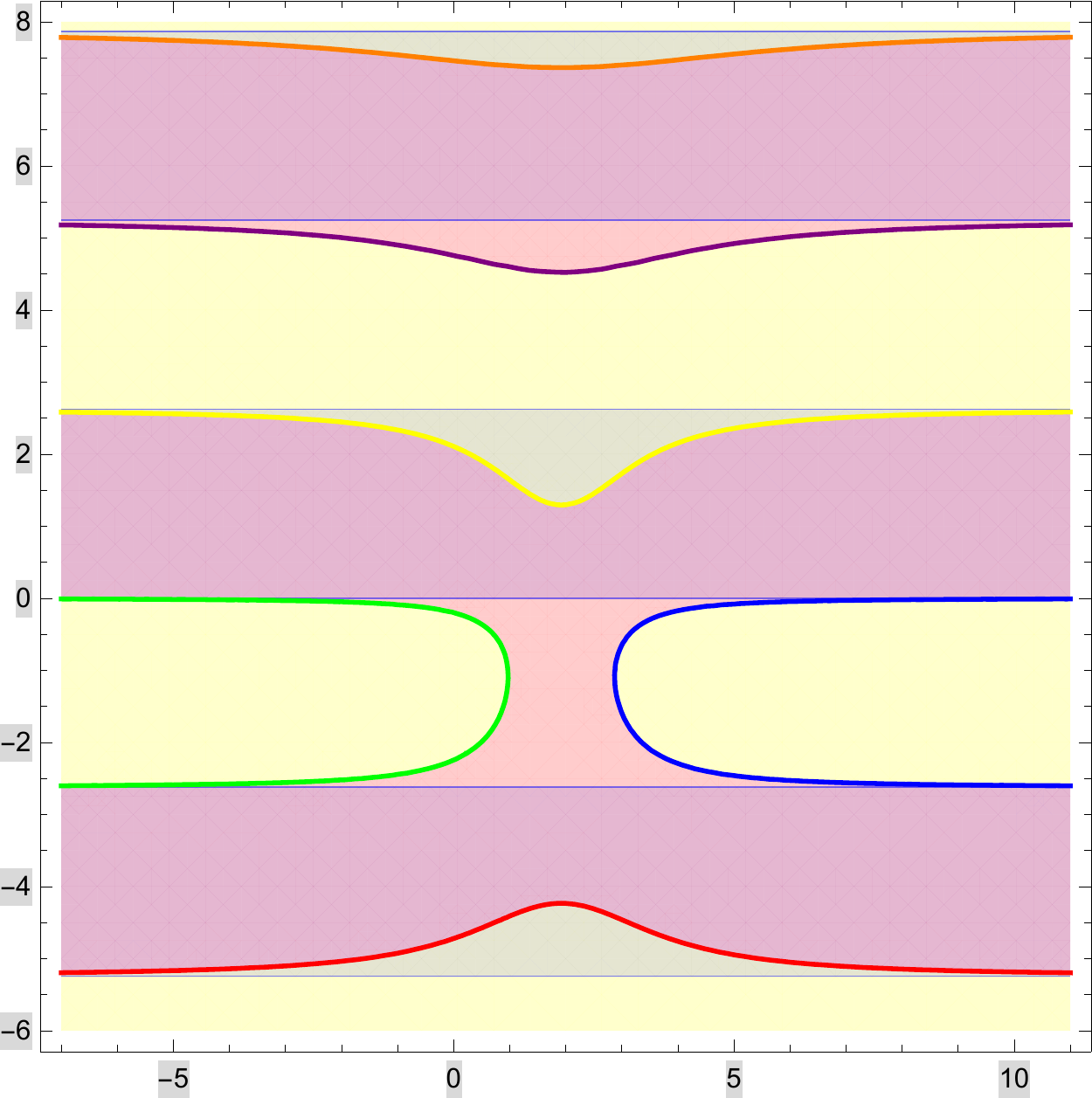}} \hspace{3mm}
    \subfloat[][]{
  \includegraphics[width=50mm]{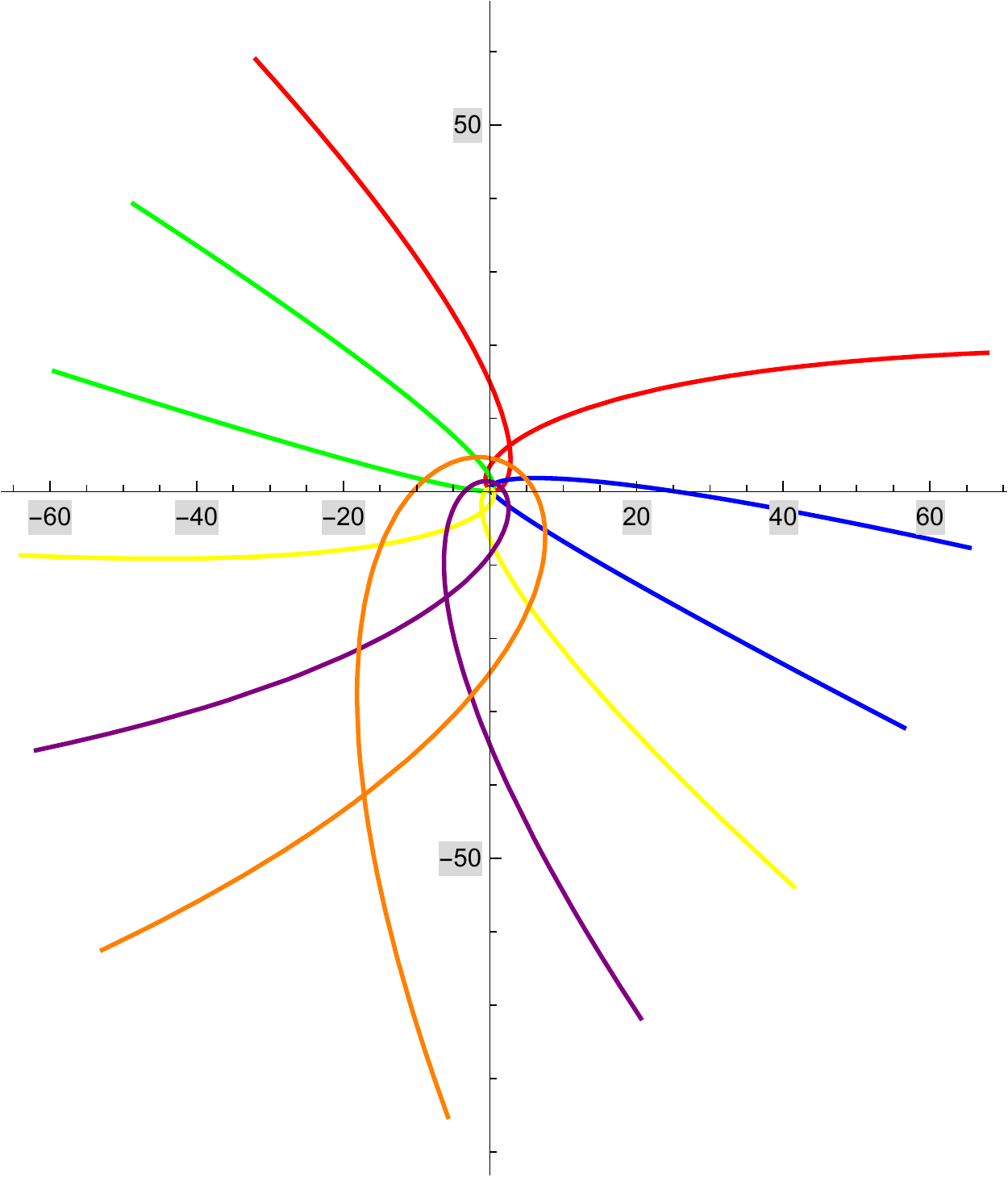}} \hspace{3mm}
    \subfloat[][]{
  \includegraphics[clip,trim=0 1.5cm 0 0 ,width=50mm]{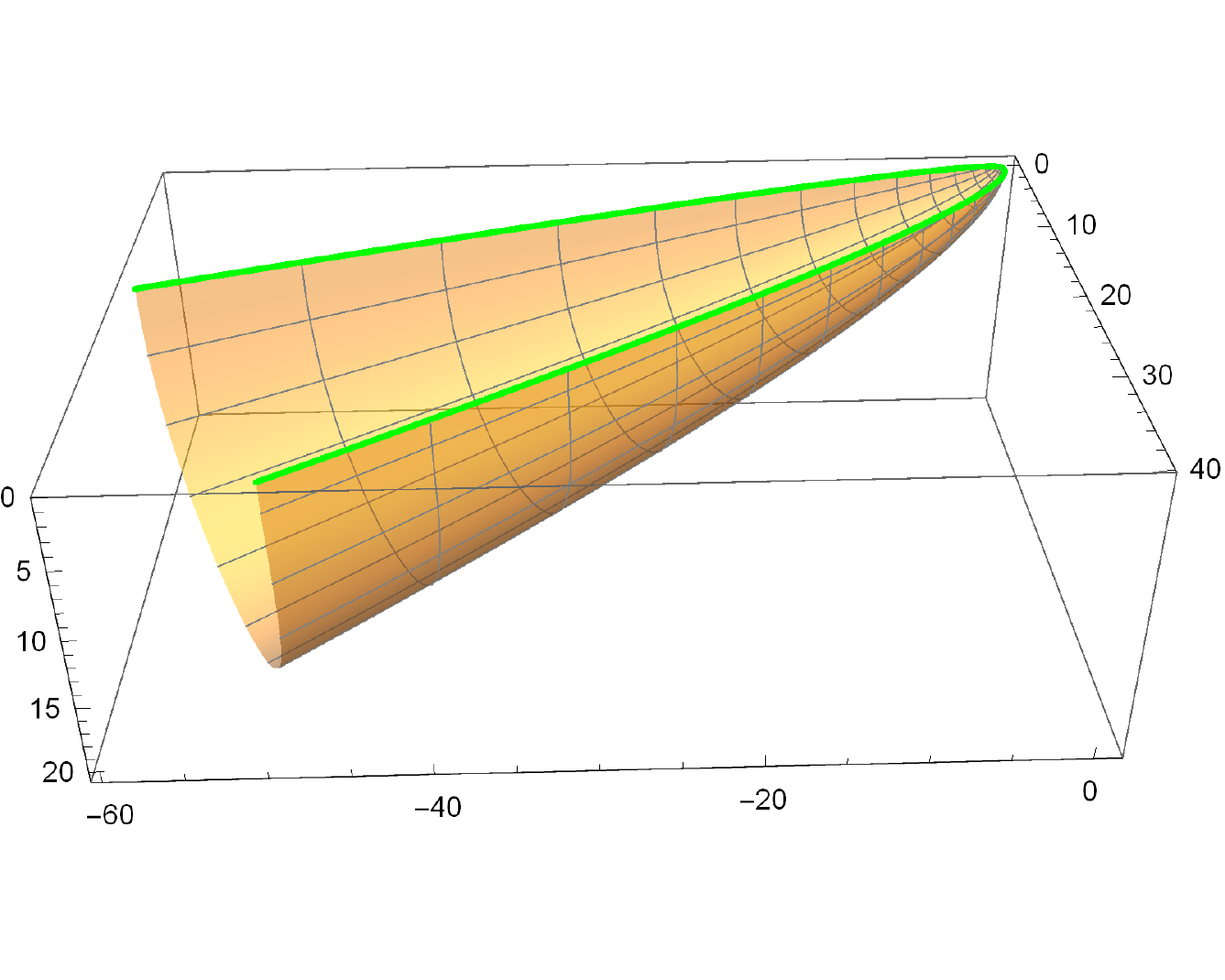}} \hspace{3mm}
    \subfloat[][]{
  \includegraphics[clip,trim=0 1.7cm 0 0 ,width=50mm]{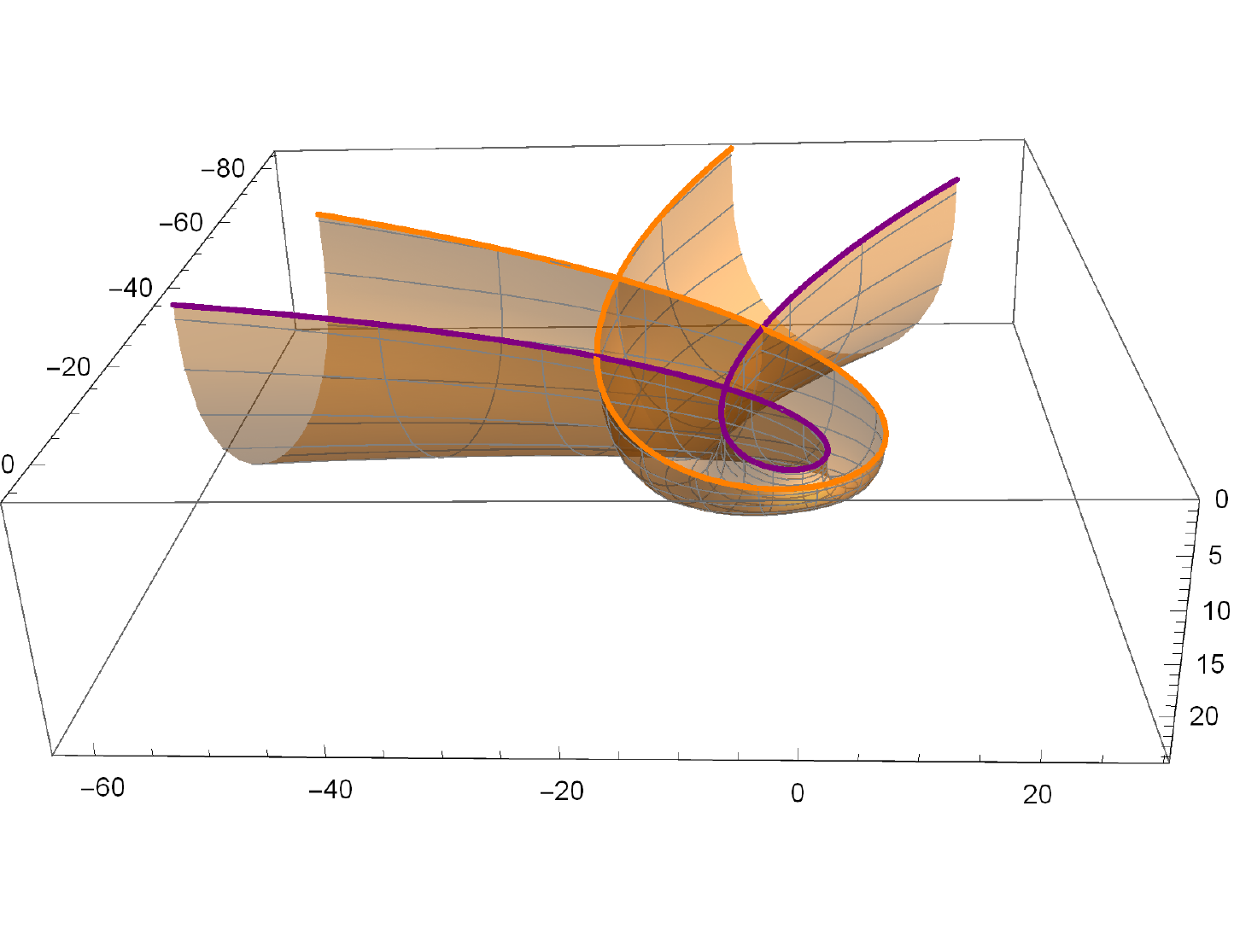}} \hspace{3mm}
    \subfloat[][]{
  \includegraphics[width=50mm]{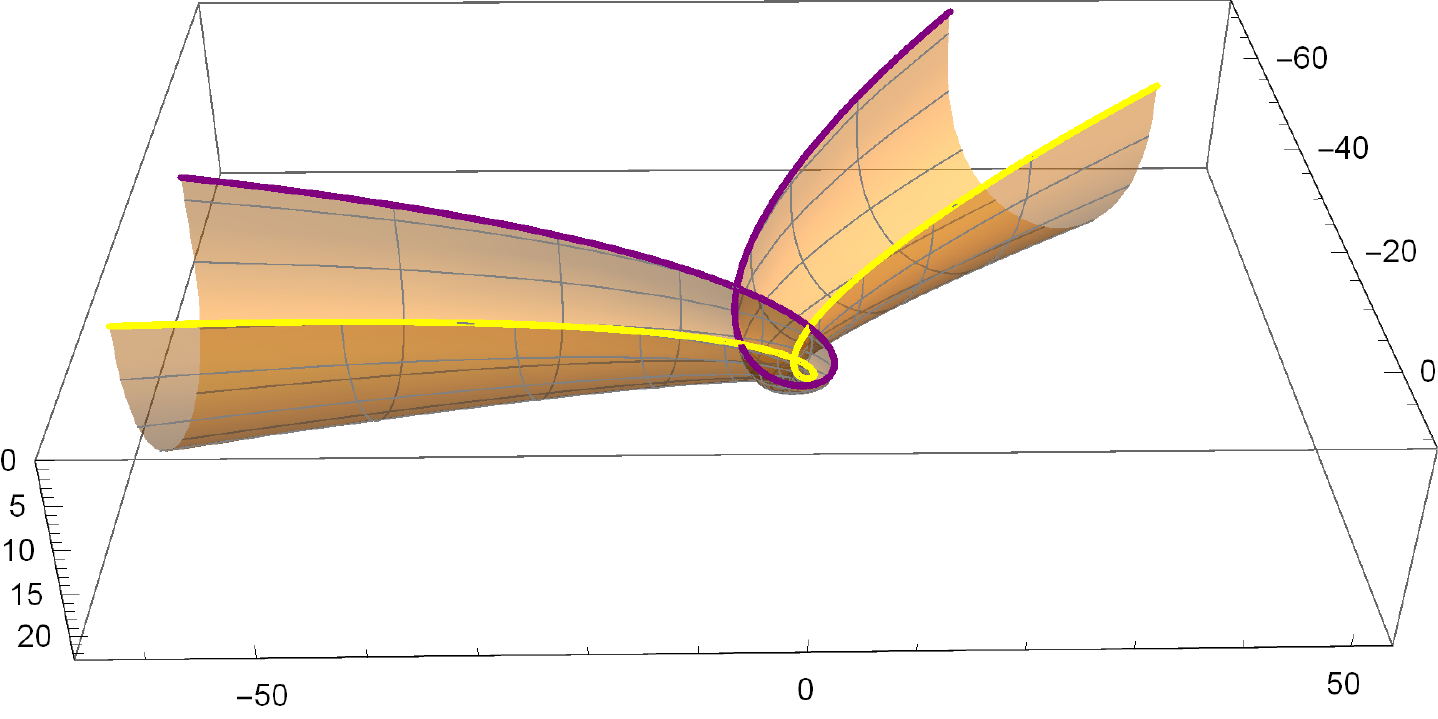}} \hspace{3mm}
    \subfloat[][]{
  \includegraphics[width=50mm]{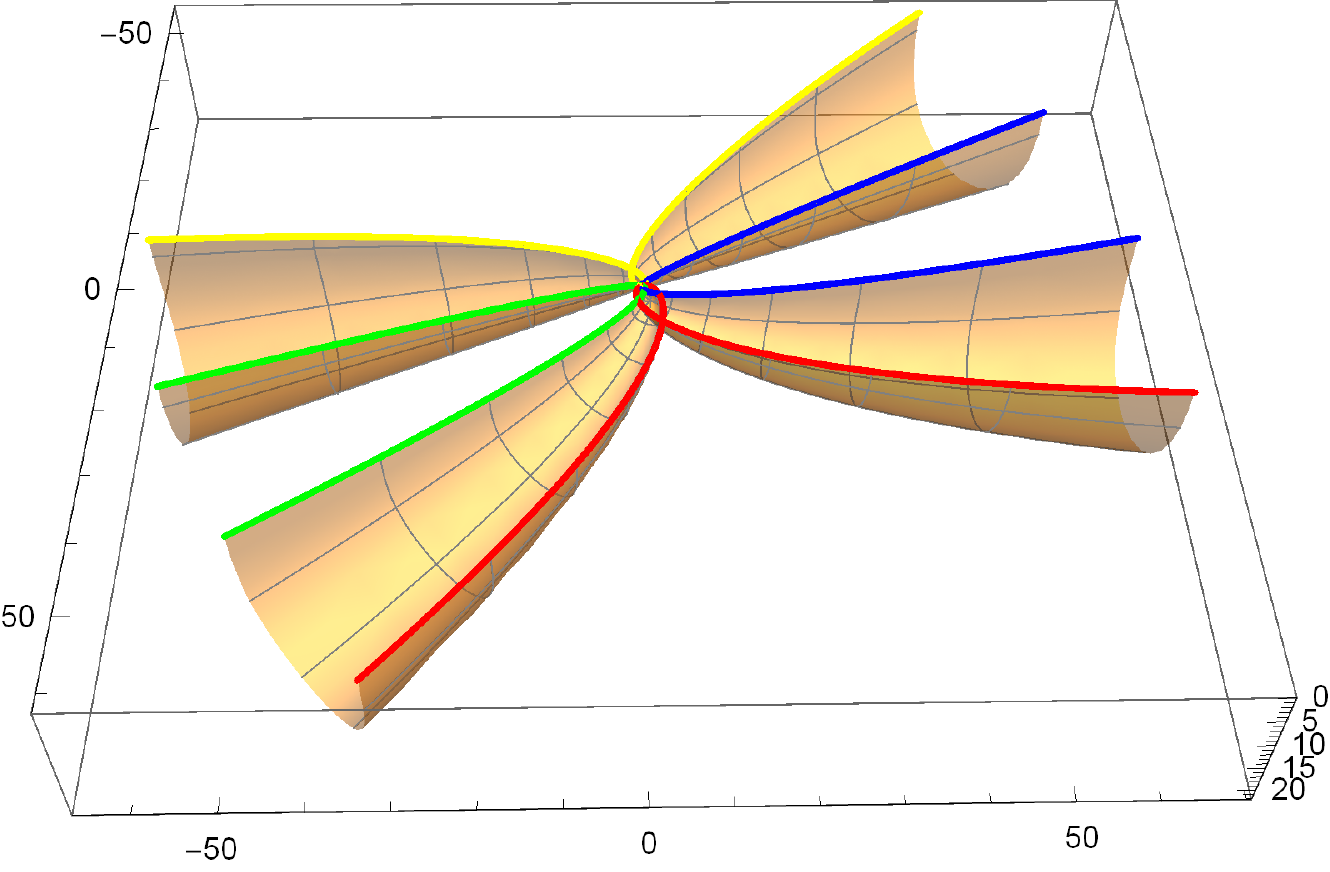}} \hspace{3mm}
    \caption{$q\bar q$ solution after the transformation with $b_1=0.1$, $b_2=0.5$.
    In (a) we plot the zeros on the worldsheet ($\hat{Z} = 0$), we denote each zero line by a different color. The red regions are mapped to $\hat{Z}>0$, and the yellow ones to $\hat{Z}<0$. The transparent blue strips represent the sections which are mapped to $Z>0$, before the transformation.
    In (b) we show how each zero line on the worldsheet is mapped to the $X_1-X_2$ plane in target space, where the colors match the ones in (a).
    In (c)-(f) we plot the minimal area surfaces in $\AdS_3$ ending on the contour in (b) denoted by the appropriate colors. All the contours except for the green and blue ones self intersect and end at infinity. In case where $\hat{Z}$ is supposed to be negative, as in (c) and (e) we simply flip the sign of $\hat{Z}$.}
    \label{fig:qqbarb05betapio22fig3new}
\end{figure}

For the original solution, $Z(\tau)=0$ for any $\tau = 2 n \EllipticK(-1)$, and the original solution (\ref{eq:qqbarsol}) is usually defined for $0<\tau<2\EllipticK(-1)$ and $-\infty<\sigma<\infty$.
However, any other range $2 n \EllipticK(-1)<\tau<2 (n+1)\EllipticK(-1)$ is also a good and equivalent solution, where for odd $n$'s we can take $Z\to -Z$.
After applying the transformation we can try to follow how each solution (namely different $n$'s) is changed as we vary $b^\mu$.
In order to the that it is convenient to look at a contour plot of the zeros of $\hat{Z}$ as in figure \ref{fig:qqbarb05betapio22fig3new} (a).
Each contour in the plot is mapped to a contour on the AdS boundary as can be seen in figure \ref{fig:qqbarb05betapio22fig3new} (b), where the colors between the figures match. The area enclosed by the contour is the minimal area surface mapped to the bulk.

This is the typical picture we get where most of the infinite strips on the worldsheet area slightly deformed on the worldsheet and the contours are mapped to self intersecting contour in target space, and the $q\bar q$-pairs continue to infinity, see figure \ref{fig:qqbarb05betapio22fig3new} (d) and (e).
Then there is the region of the worldsheet enclosed by the yellow and red contours in figure \ref{fig:qqbarb05betapio22fig3new} (a), which gives two different types of Wilson loops, the self-intersecting one in \ref{fig:qqbarb05betapio22fig3new} (f), and the smooth non-self intersecting contours (green and blue), as in \ref{fig:qqbarb05betapio22fig3new} (c).
This picture changes with $b^\mu = b(\cos\beta,\sin\beta)$ as follows, starting with $\beta=0$, the nonintersecting contours (blue and green in the figure) appear in lower $\tau$ strips as $b$ decreases, and the other way when $b$ increase up to a critical value $b =\frac{1}{2(\EllipticE-\EllipticK)} \simeq 0.8346..$ from which these contours appear in the $n=-1$ strip (where $\EllipticK \equiv \EllipticK(-1)$ and $\EllipticE \equiv \EllipticE(-1)$).
As we increase $\beta$ these contours move to higher $\tau$ strips until the picture gets reflected with respect to the $\sigma$ axis when $\beta = \pi$. For $\beta = \pi/2$ these two contours meet at the origin of the worldsheet and the worldsheet picture is symmetric with respect to the $\sigma$ axis.

The new solutions seem to be genus-1 solutions by the appearance of the elliptic functions.
If this is indeed the case, it would be interesting show it explicitly and find a relation between the $b^\mu$ parameter and the algebraic curve parameter $a$, where $y^2 = \lambda(\lambda-a)(\lambda-1/\bar a)$ is the algebraic curve equation.
The general genus-1 solution is given in \cite{Kruczenski:2013bsa}, though similar solution are not described there, see also \cite{Cooke:2014uga}.

\subsubsection*{Special cases}

\begin{figure}
    \centering
  \includegraphics[clip, trim=0 2.8cm 0 2.8cm ,width=50mm]{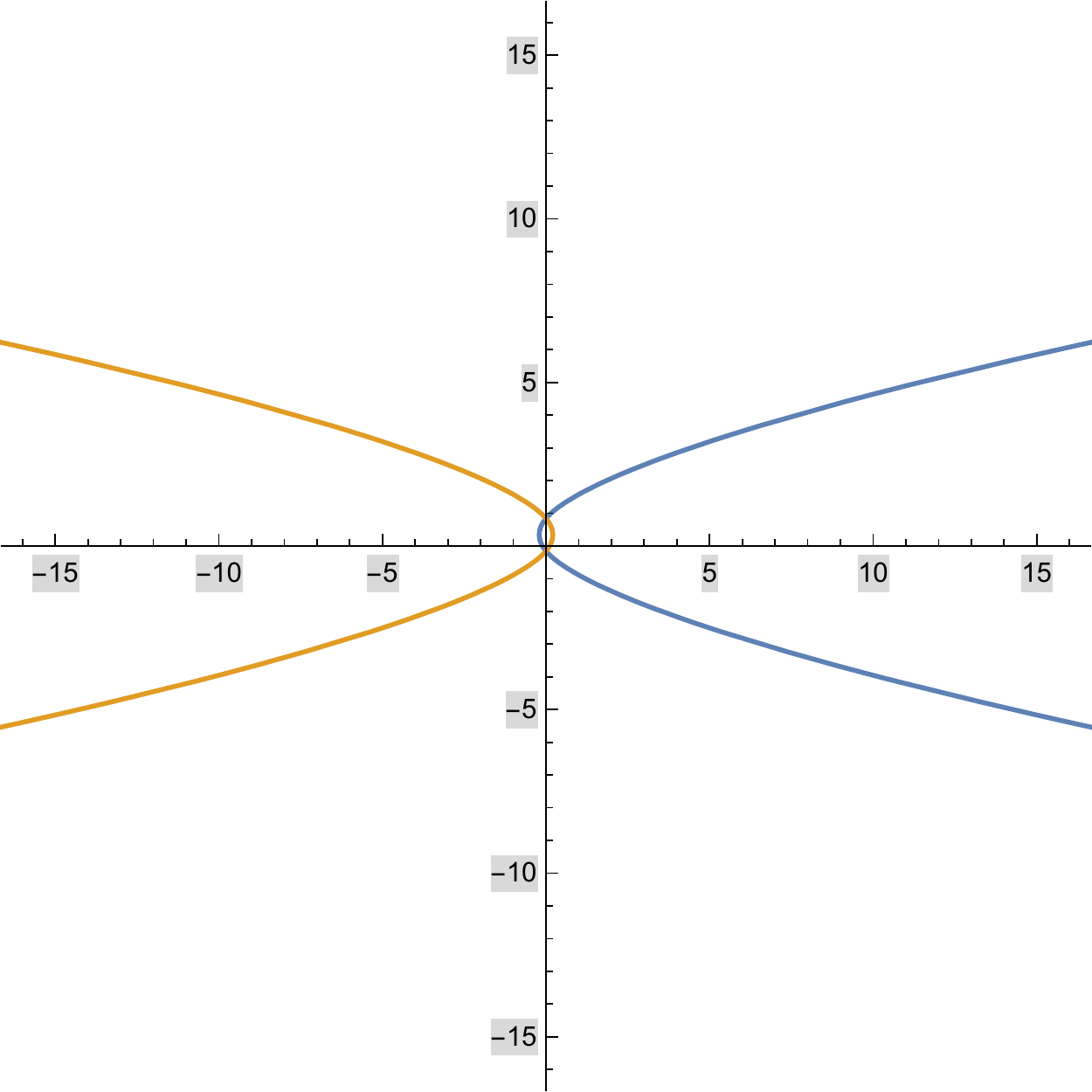} \hspace{3mm}
    \caption{Contours of two symmetric Wilson loops on the AdS boundary ($X_1-X_2$ plane), generated from the $q\bar q$ potential solution. We can consider each one of them (blue and orange) separately, similar to \ref{fig:qqbarb05betapio22fig3new} (c).}
    \label{fig:qqbarspecial}
\end{figure}

There is at least one special choice of the parameters which results in a symmetric contour in target space.
For $b_2 = 0$ and $b_1 = \frac{1}{3(\EllipticE-\EllipticK)} \simeq 0.55641789$, the two contours defined by
\begin{align}
\sigma = \pm \sqrt{2(\tilde\EllipticE - 3(\EllipticE-\EllipticK))\JacobiCN\JacobiDS-(\tilde\EllipticE - 3(\EllipticE-\EllipticK))^2+\JacobiSN^2},
\end{align}
where $-4 \EllipticK<\tau<- 2 \EllipticK$, are symmetric with respect to the line $X_2 = \frac{72 \pi ^5-\Gamma \left(\frac{1}{4}\right)^8+12 \pi ^2 \Gamma \left(\frac{1}{4}\right)^4}{36 \sqrt{2} \pi ^{7/2} \Gamma \left(\frac{1}{4}\right)^2} \simeq 0.343697$, see figure \ref{fig:qqbarspecial}.
Asymptotically for large $X_1$ we have $X_2 \simeq \pm X_1^{2/3}(\EllipticE-\EllipticK)^{1/3}+\mathcal{O}(X_1^0)$.

When $b\to \infty$ the AdS boundary nonintersecting contour is composed from two curves, where one one them is straight.

\subsection{$\left<W \mathcal{O}_{\mathcal{J}}\right>$}

\begin{figure}
    \centering
    \subfloat[][]{
  \includegraphics[width=50mm]{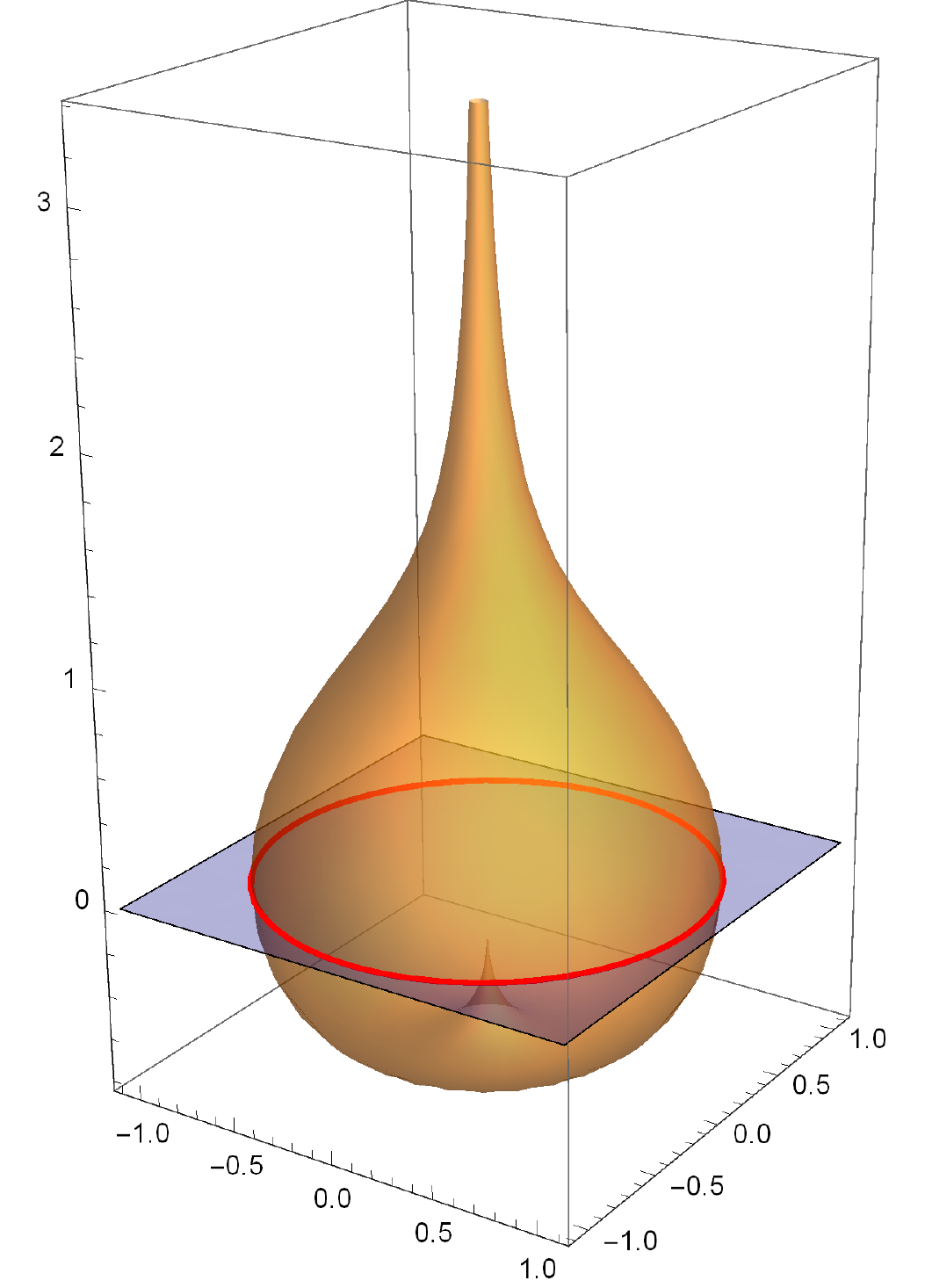}} \hspace{3mm}
    \subfloat[][]{
  \includegraphics[width=50mm]{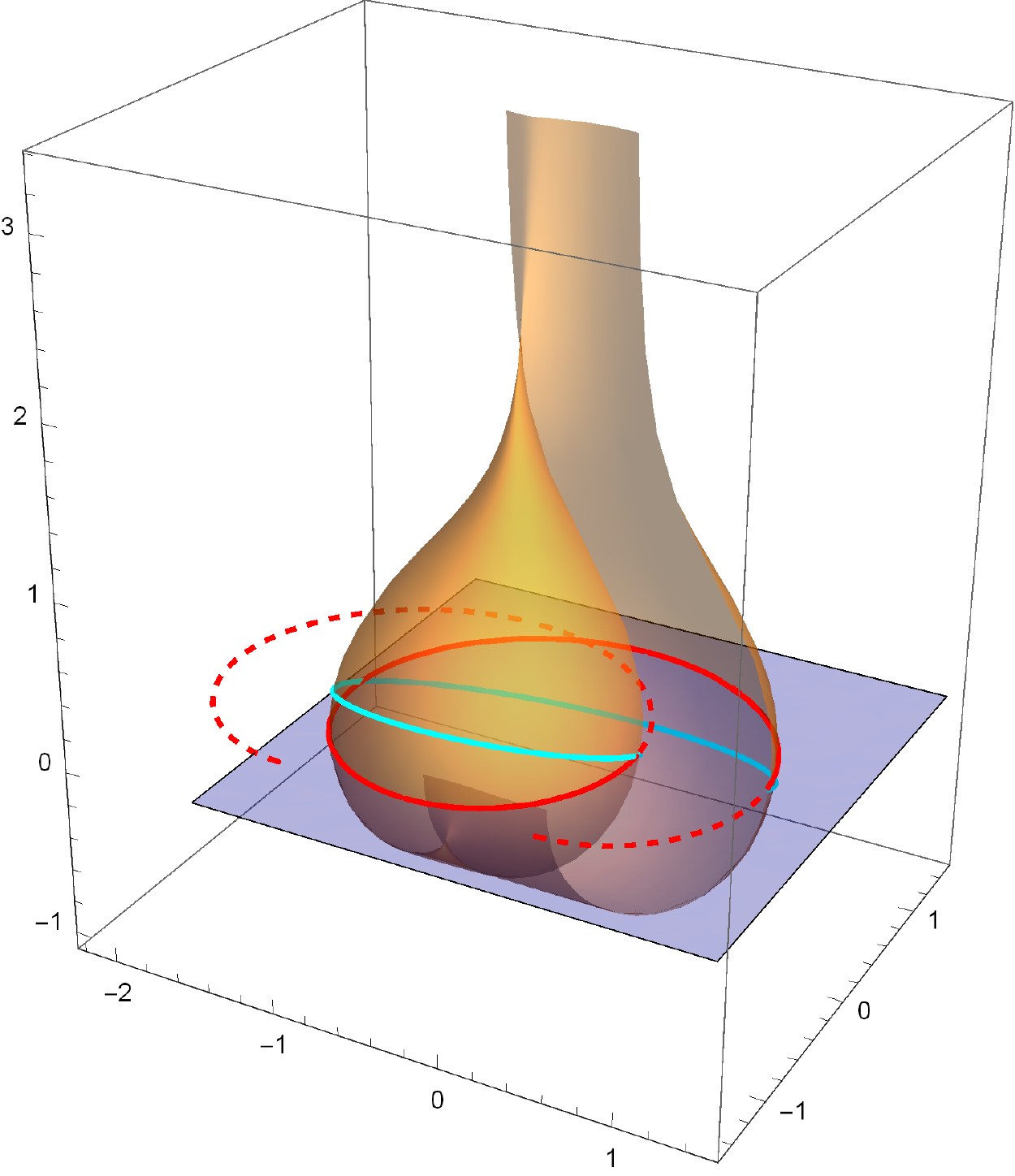}} \hspace{3mm}
\subfloat[][]{
  \includegraphics[width=50mm]{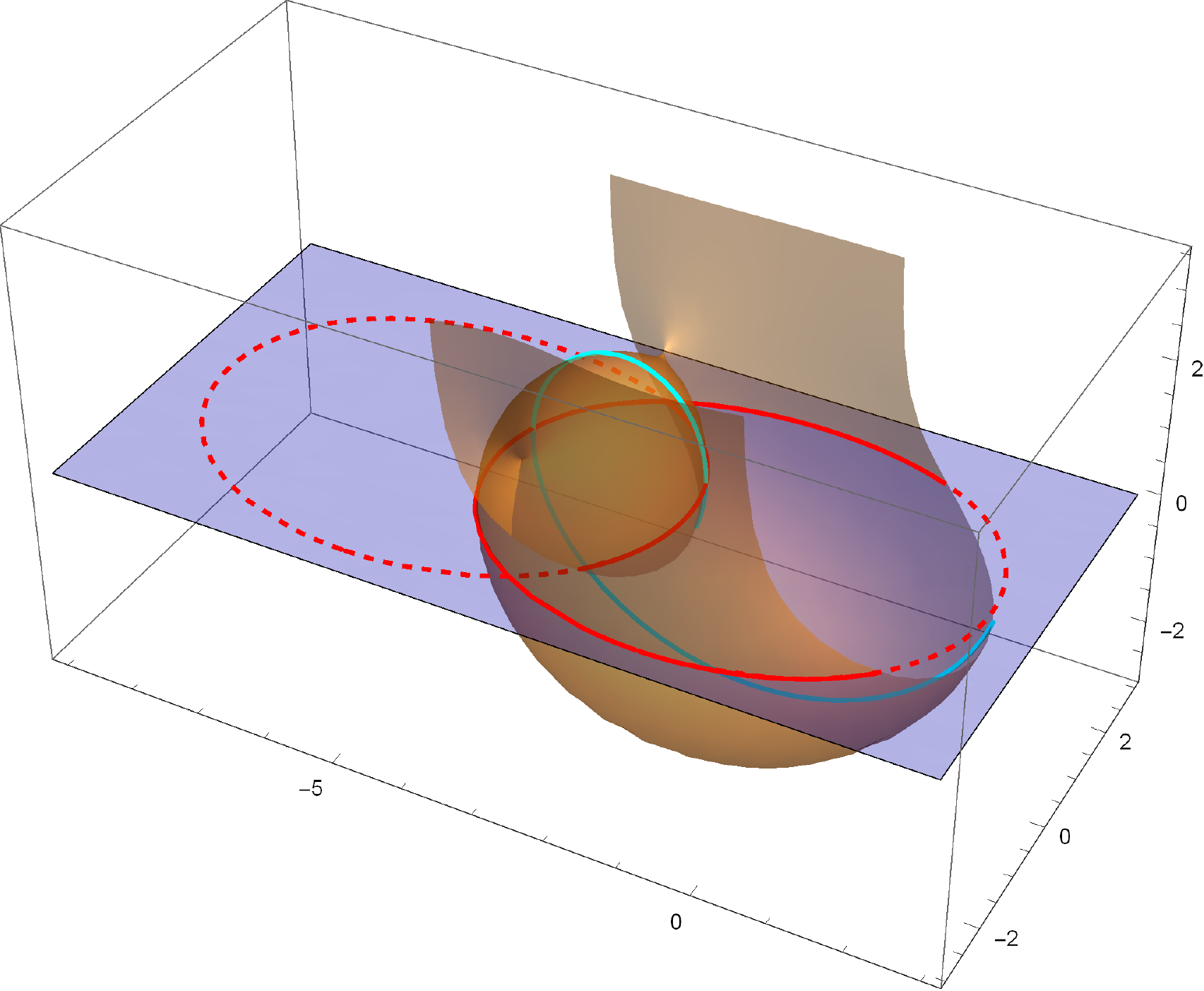}} \hspace{3mm}
   \caption{The $\left<W \mathcal{O}_{\mathcal{J}}\right>$ surface in $\AdS_3$ for $\mathcal{J} = 0.3$, $-\infty < \tau <\infty$, $0<\sigma<2\pi$.
   The blue surface represents the $\AdS$ boundary ($Z=0$) and the red contour is the boundary contour created by the new surface, the dashed part corresponds to extension of the $\sigma$ worldsheet coordinate.
   The "old" boundary contour is mapped to the cyan colored line, this curve intersects the boundary and the new boundary contour twice.
   We plotted the surface also for negative values of $Z$, since the non-local transformation acts on the whole surface regardless if it is above or below the boundary.
   (a) The original solution which is periodic in $\sigma$ and has two spikes, one below the boundary which ends on it, and another above the boundary which ends at $z\to \infty$. This corresponds to $b=0$. For $\mathcal{J} \to 0$ we recover the circular Wilson loop with no operator.
   (b) $b=0.2$, when we start to increase $b$ the original surface "opens" and is no longer periodic in $\sigma$. The boundary curve becomes infinite periodic self intersecting curve.
   (c) $b=1.06176..$, at a critical value of $b$ (which is $b=0.52922$ in this case) smooth closed contours start to form. This pattern exists for a continuous range of b, which is $0.52922<b<1.59429$ is this case. We can isolate these contours and make sense out of them as smooth Wilson loops.
   In this case, the $\Phi \in \Sphere^1$ coordinate is a smooth imaginary function of the contour parameter, $\gamma$, with on minimum and one maximum (remember that $\Phi = i \mathcal{J} \tau$).
    }
    \label{fig:WOfig}
\end{figure}

After exploring some solutions in $\AdS_3$, we continue to higher dimensions.
The first solution we study corresponds to the correlation function of a circular Wilson loop and and a BMN operator, $\left<W \mathcal{O}_{\mathcal{J}}\right>$, which is 1/4-BPS.
The original solution is given by \cite{Zarembo:2002ph}
\begin{align}
X_1 + i X_2= \frac{\sqrt{\mathcal{J}^2+1} e^{\mathcal{J} \tau }e^{i\sigma}}{\cosh \left(\sqrt{\mathcal{J}^2+1} (\tau +\tau_0)\right)},\quad
Z   = \frac{e^{\mathcal{J} \tau } \sinh \left(\sqrt{\mathcal{J}^2+1} \tau \right)}{\cosh \left(\sqrt{\mathcal{J}^2+1} (\tau +\tau_0)\right)},\quad
\Phi = i \mathcal{J} \tau,
\end{align}
with $\tau_0 = \frac{\sinh ^{-1}(\mathcal{J})}{\sqrt{\mathcal{J}^2+1}}$, where $0\leq \tau$ and $0\leq \sigma < 2\pi$.
It is a surface of revolution in $\AdS_3$, periodic in $\sigma$, which lives in $\AdS_3 \times \Sphere^1$, see figure \ref{fig:WOfig} (a).

After the non-local transformation, we end up with a string solution which generally self intersects and is no longer periodic in $\sigma$, see figure \ref{fig:WOfig} (b).
The new solution is given by
\begin{align}
\hat{X}_1 + i \hat{X}_2 & ~=
\frac{J e^{\mathcal{J} \tau +i \sigma }}{J \text{ch} +\mathcal{J} \text{sh}}
-\frac{2 b e^{i \beta } \left(\mathcal{J} \sigma  (J \text{ch}+\mathcal{J} \text{sh})-i \left(J \mathcal{J} \text{ch} +J^2 \text{sh}\right)\right)}{J \text{ch} +\mathcal{J} \text{sh}}
+\frac{b^2 J e^{2 i \beta -\mathcal{J} \tau -i \sigma }}{J \text{ch} +\mathcal{J} \text{sh}},\nonumber\\
\hat{Z} & ~=
\frac{e^{\mathcal{J} \tau }\text{sh} }{J \text{ch} +\mathcal{J} \text{sh}}
+\frac{2 b J \sin (\beta -\sigma )}{J \text{ch}+\mathcal{J} \text{sh}}
-\frac{b^2 e^{-\mathcal{J} \tau } \left(2 J \mathcal{J} \text{ch}+(2 \mathcal{J}^2 +1) \text{sh}\right)}{J\text{ch}+\mathcal{J} \text{sh}}
,\quad
\Phi = i \mathcal{J} \tau,
\end{align}
where we defined $J\equiv \sqrt{\mathcal{J}^2 +1}$ and $\text{ch} \equiv \cosh(\sqrt{\mathcal{J}^2 +1} \tau)$ and $\text{sh} \equiv \sinh(\sqrt{\mathcal{J}^2 +1} \tau)$.

Obviously the transformation does not correspond to a conformal transformation of the initial solution.
Generally, when we increase $b$ from zero, the original surface "opens" and the boundary curve becomes an infinite periodic smooth self-intersecting curve, where we extend the range of $\tau$ and $\sigma$.
However, for a certain range of values $b_{\text{min}}(\mathcal{J})<b<b_{\text{max}}(\mathcal{J})$ infinitely many smooth closed contours start to form.
Each closed contour can be identified as a smooth Wilson loop with non-trivial function on $\Sphere^1$.
The contours are also closed on the worldsheet, so the angle on the sphere has a minimum value which increases as we go along the loop until it reaches a maximum value from which it decreases until we get back to the minimum point.

\begin{figure}
    \centering
\subfloat[][]{
  \includegraphics[width=35mm]{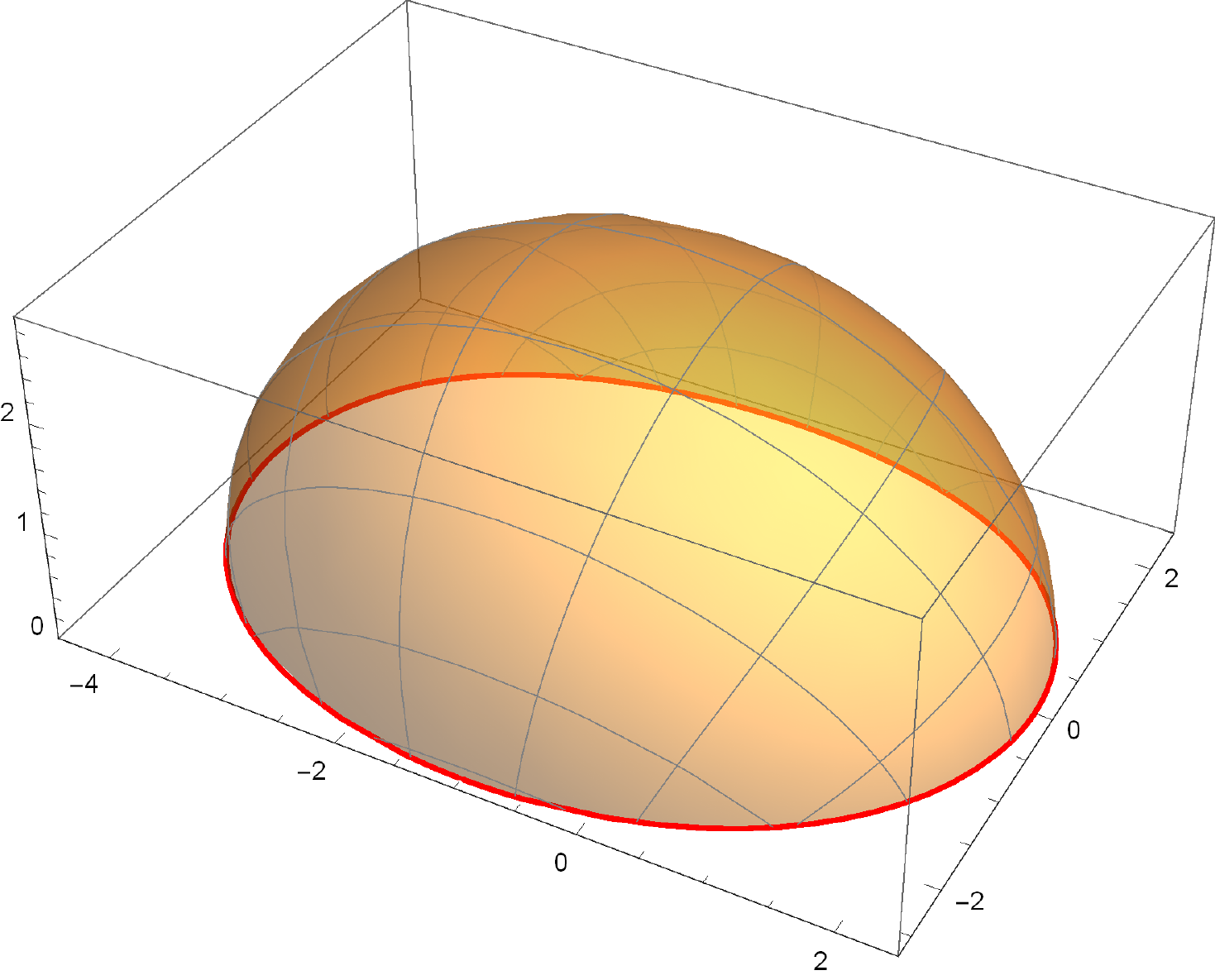}} \hspace{2mm}
\subfloat[][]{
  \includegraphics[width=35mm]{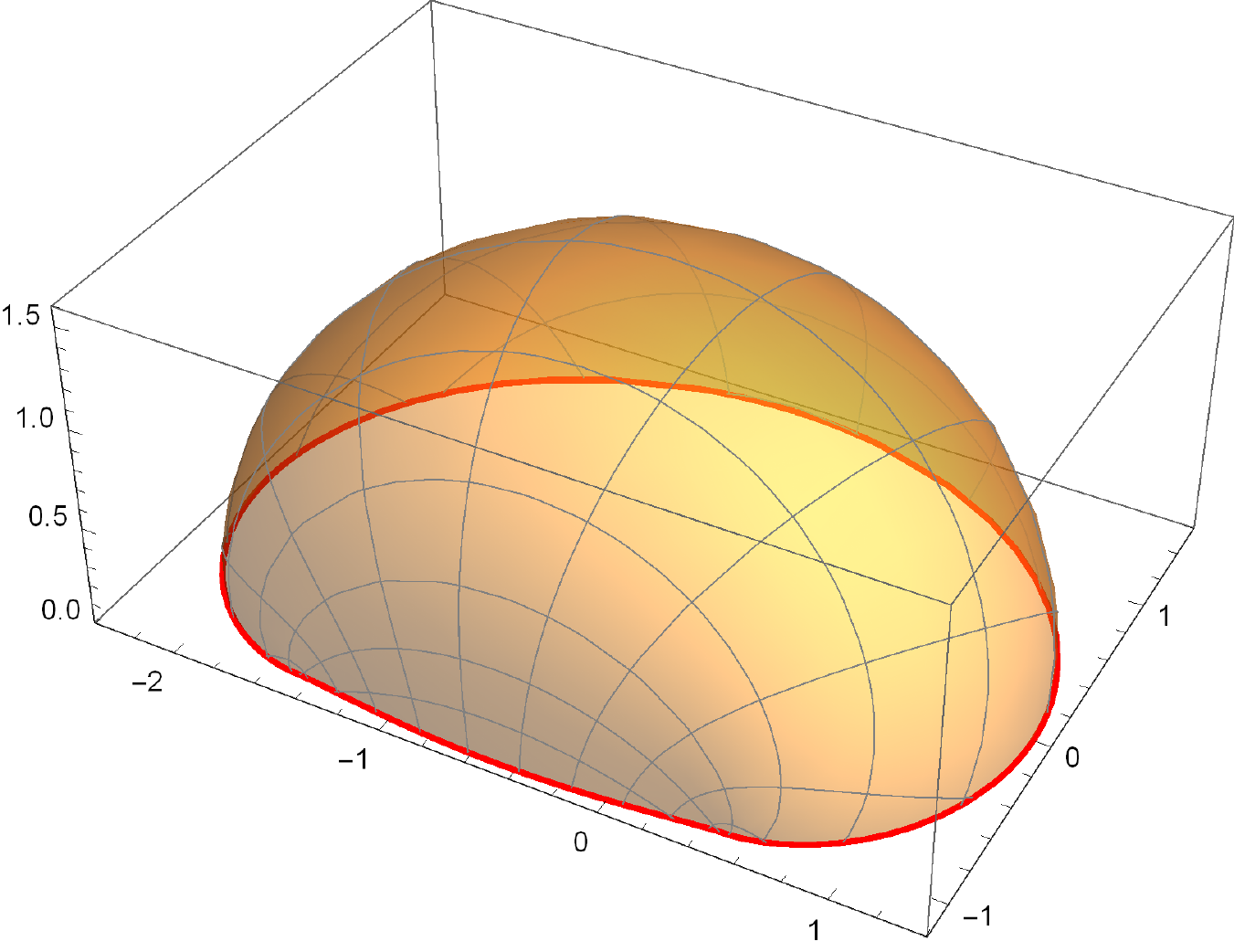}} \hspace{2mm}
      \subfloat[][]{
  \includegraphics[width=35mm]{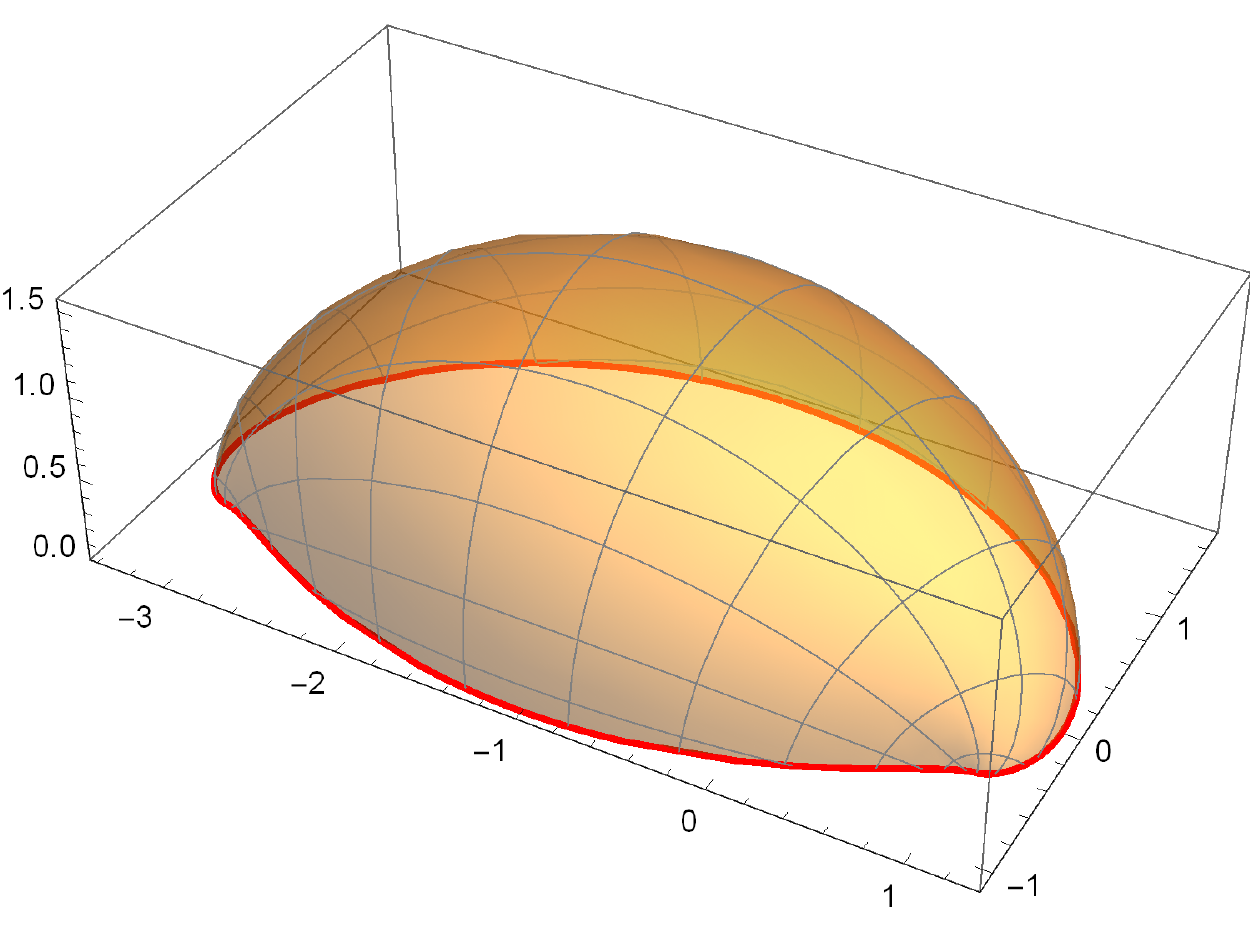}} \hspace{2mm}
\subfloat[][]{
  \includegraphics[width=35mm]{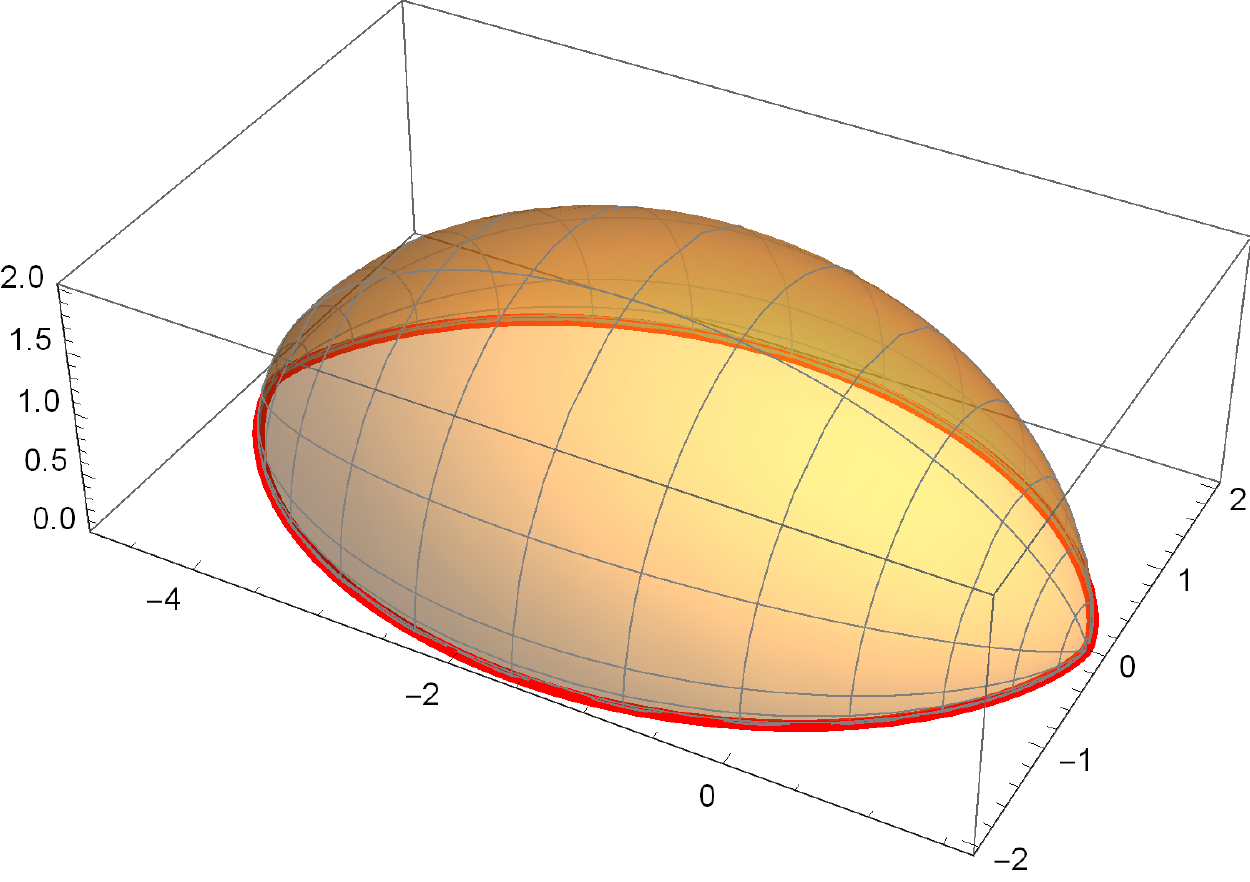}} \hspace{2mm}
   \caption{Some examples of the resulting minimal area surfaces Wilson loops plotted in $\{X_1,X_2,Z\}$ space, for
   (a) $\mathcal{J}=0.3$, $b=1.06$.
   (b) $\mathcal{J}=0.3$, $b=0.529$,
   (c) $\mathcal{J}=1.5$, $b=0.22$,
   (d) $\mathcal{J}=5000$, $b=\frac{1}{2\mathcal{J}}$. In (d) the boundary contour is approximated by the $\mathcal{J}\to \infty$ contour which is composed of two cycloids. In (b) and (c) $b$ takes is the minimal value as a function of $\mathcal{J}$ to form a closed contour.
   The red curves are the boundary contours.
    }
    \label{fig:WOsurface}
\end{figure}

\begin{figure}
    \centering
    \includegraphics[trim = 0mm 0mm 0mm 0mm,clip,width = 0.6\textwidth]{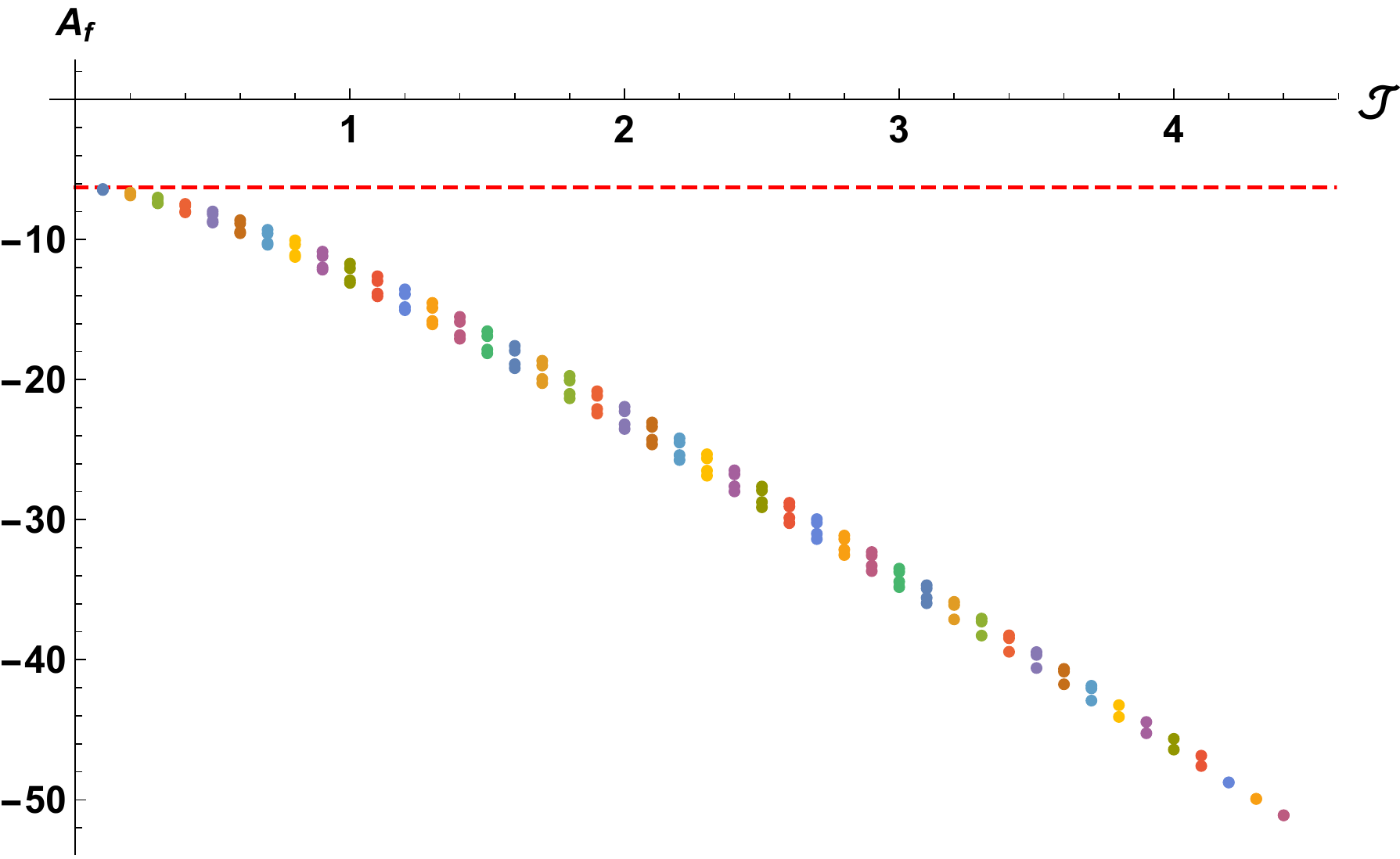}
    \caption{Numerical evaluation of the Wilson loops area as a function of $\mathcal{J}$.
    For each $\mathcal{J}$ we evaluate the area for several values of $b$ for which the Wilson loops is closed. The red line indicates the circular Wilson loop bound, namely $-2\pi$.}
    \label{fig:WOarea}
\end{figure}

Concentrating on these smooth closed Wilson loops, we can express their boundary curve analytically, though we evaluate their regularized area numerically. Some examples of such Wilson loops are given in figure \ref{fig:WOsurface}.
As explained above, for any $\mathcal{J}$ there is a range of values of $b$ which correspond to a smooth closed contour.
Different values of $b$ correspond to Wilson loops with different expectation values, and different $\Phi$ functions.
We plot the area of these Wilson loops for $0<\mathcal{J}<4.5$ where we sample several relevant $b$ values for each $\mathcal{J}$, see figure \ref{fig:WOarea}. Clearly the area decrease with $\mathcal{J}$ growing, and it asymptotes to the circular Wilson loops value for $\mathcal{J} \to 0$ as expected.

For large $\mathcal{J}$ we observe that choosing $b \sim \frac{1}{2 \mathcal{J}}$ results in smooth closed Wilson loops.
In order to take the limit one should rescale $\tau$ by $\mathcal{J}^{-1}$ and then take the large $\mathcal{J}$ limit.
The resulting contour is composed of two cycloids in the $\mathcal{J} \to \infty$ limit, see figure \ref{fig:WOsurface} (d).
In the $b\to \infty$ limit the solution transforms back to itself (up to a conformal transformation and reparametrization).

\subsection{The longitude}

Next we consider the 1/4 BPS longitude solution living in $\AdS_4 \times \Sphere^2$ \cite{Drukker:2007qr}.
The AdS part of the solution is given by
\begin{align}
& X^\mu = \left(\frac{a \sin \sigma  \sin a \sigma +\cos \sigma \cos a \sigma }{\cosh \sqrt{1-a^2} \tau },
\frac{a \sin \sigma  \cos a \sigma  -\cos \sigma  \sin a \sigma }{\cosh \sqrt{1-a^2} \tau },
-\tanh \sqrt{1-a^2} \tau \right),\nonumber\\
& Z = \frac{\sqrt{1-a^2} \sin \sigma }{\cosh \sqrt{1-a^2} \tau },
\end{align}
so the energy momentum tensor on the AdS part is given by $T = a^2$.
$a$ is related to the opening angle $\delta$ by, $a = \frac{\pi - \delta}{\pi}$, so $|a| \leq 1$.
Acting with the dual special conformal transformation we get a new solution with three continuous parameters\footnote{We could have considered the solution in $\AdS_5$ and have four transformation parameters, but for simplicity we only consider transformations in the $\AdS_4$ subspace.} corresponding to the vector $b^\mu = (b_1, b_2, b_3) \equiv b(\cos\beta\cos B, \cos\beta\sin B, \sin\beta)$.
The resulting expression for the new solution in quite lengthy, in the following we shall consider some special choices for the parameters which yields simple expressions. The full solution is given in appendix \ref{ap:long}.

\subsubsection*{$b = \sqrt{1-a^2},\beta = B =0$}
\begin{figure}
    \centering
    \subfloat[][]{
  \includegraphics[width=40mm]{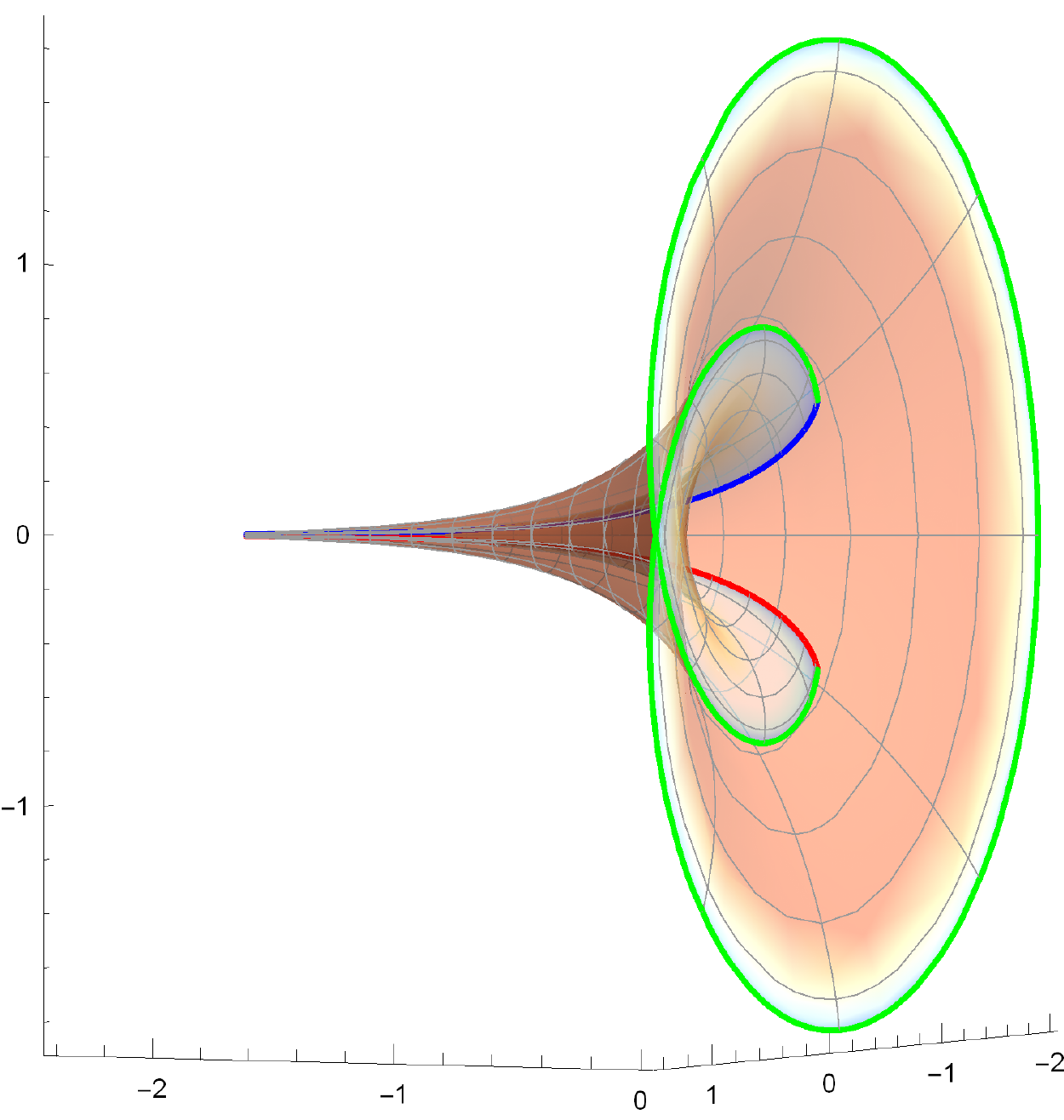}} \hspace{3mm}
    \subfloat[][]{
  \includegraphics[width=40mm]{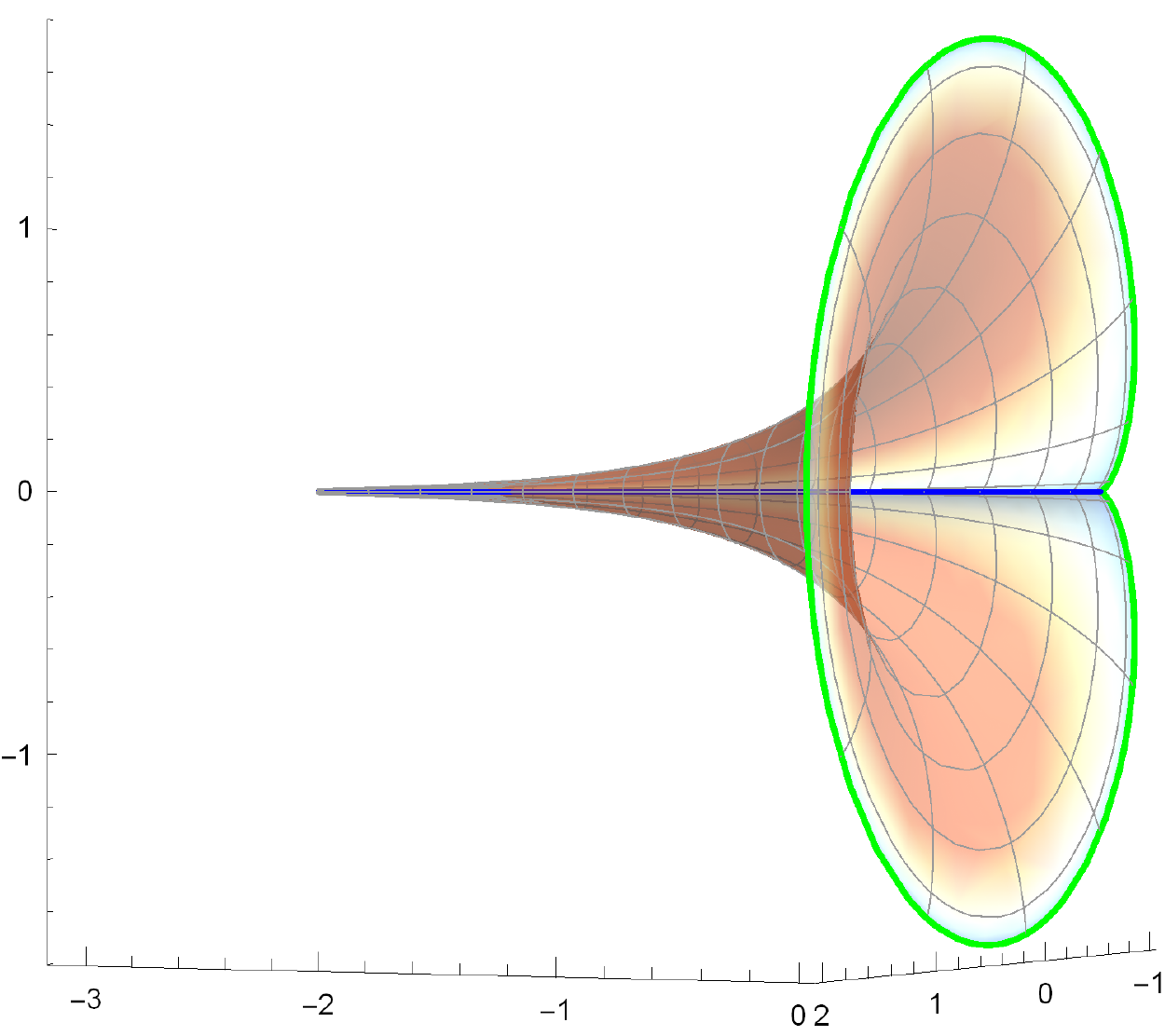}} \hspace{3mm}
    \subfloat[][]{
  \includegraphics[width=40mm]{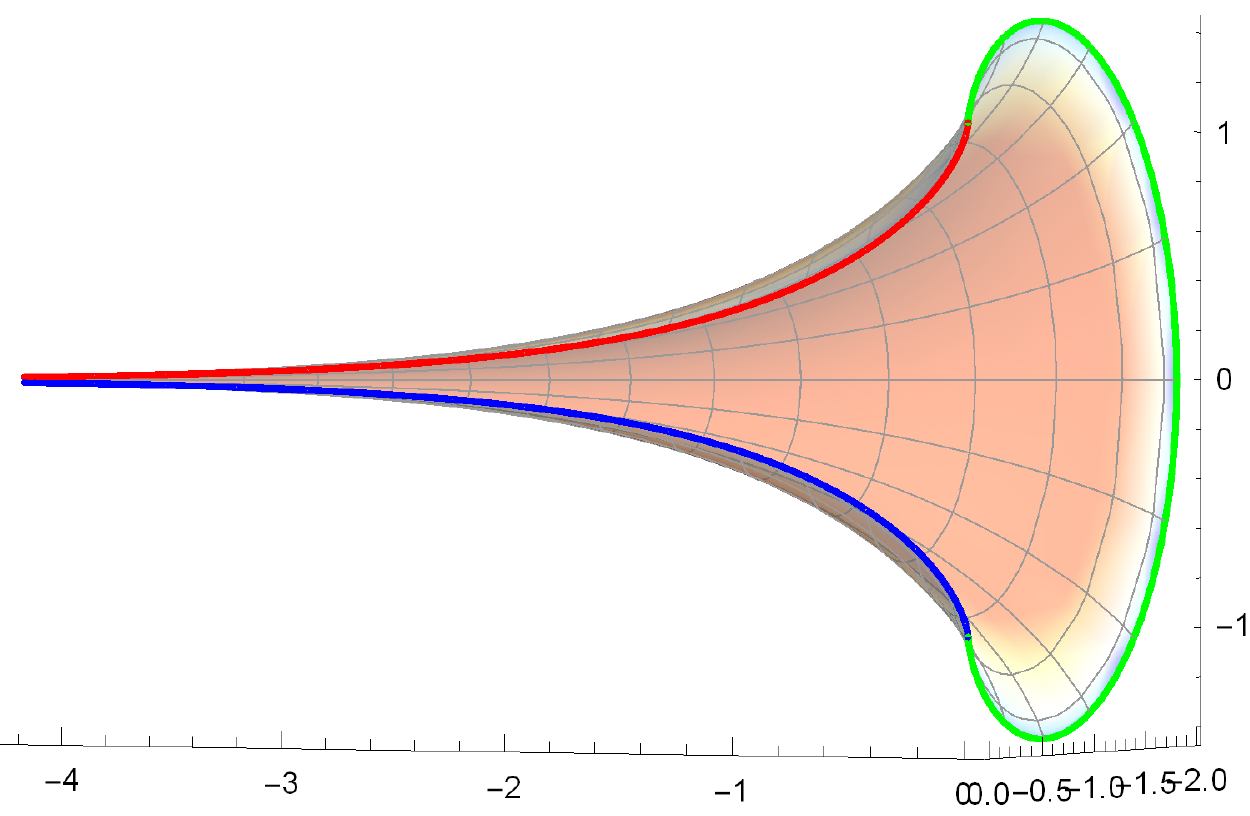}} \hspace{3mm}
      \subfloat[][]{
  \includegraphics[width=40mm]{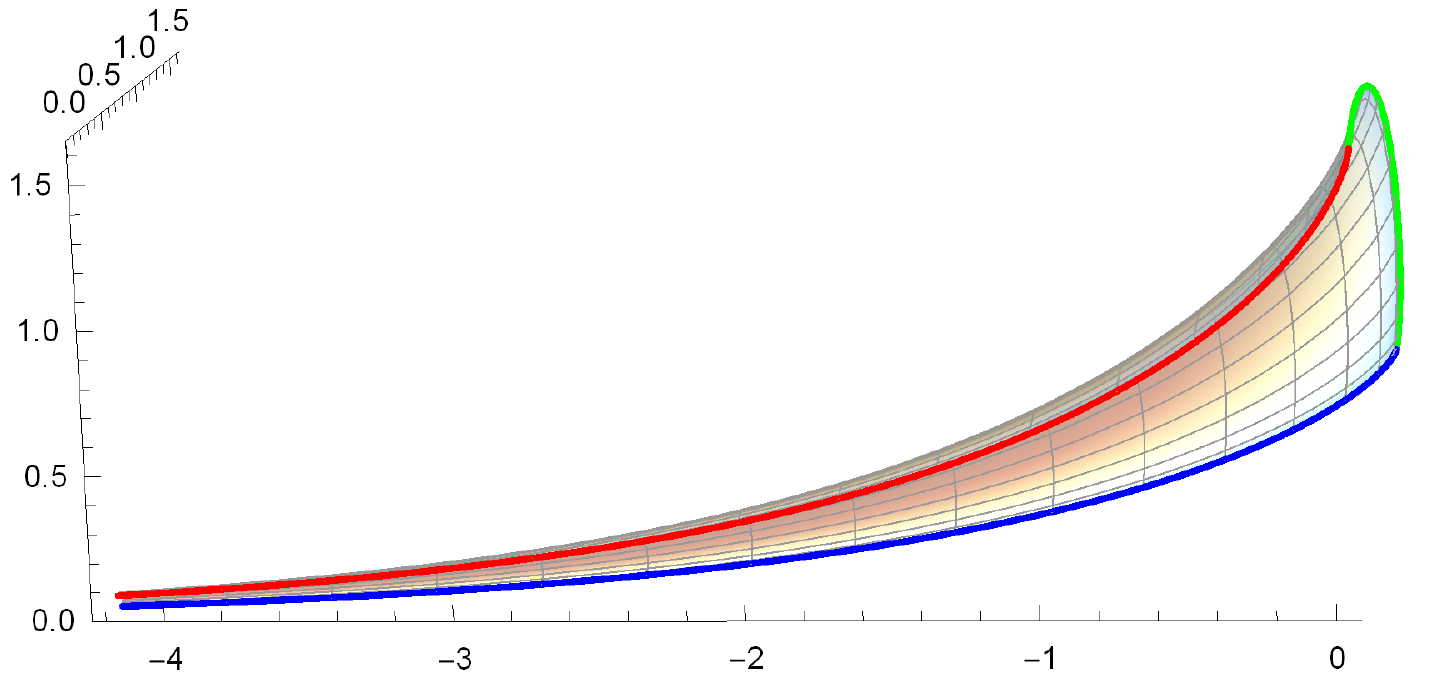}} \hspace{3mm}
      \subfloat[][]{
  \includegraphics[width=40mm]{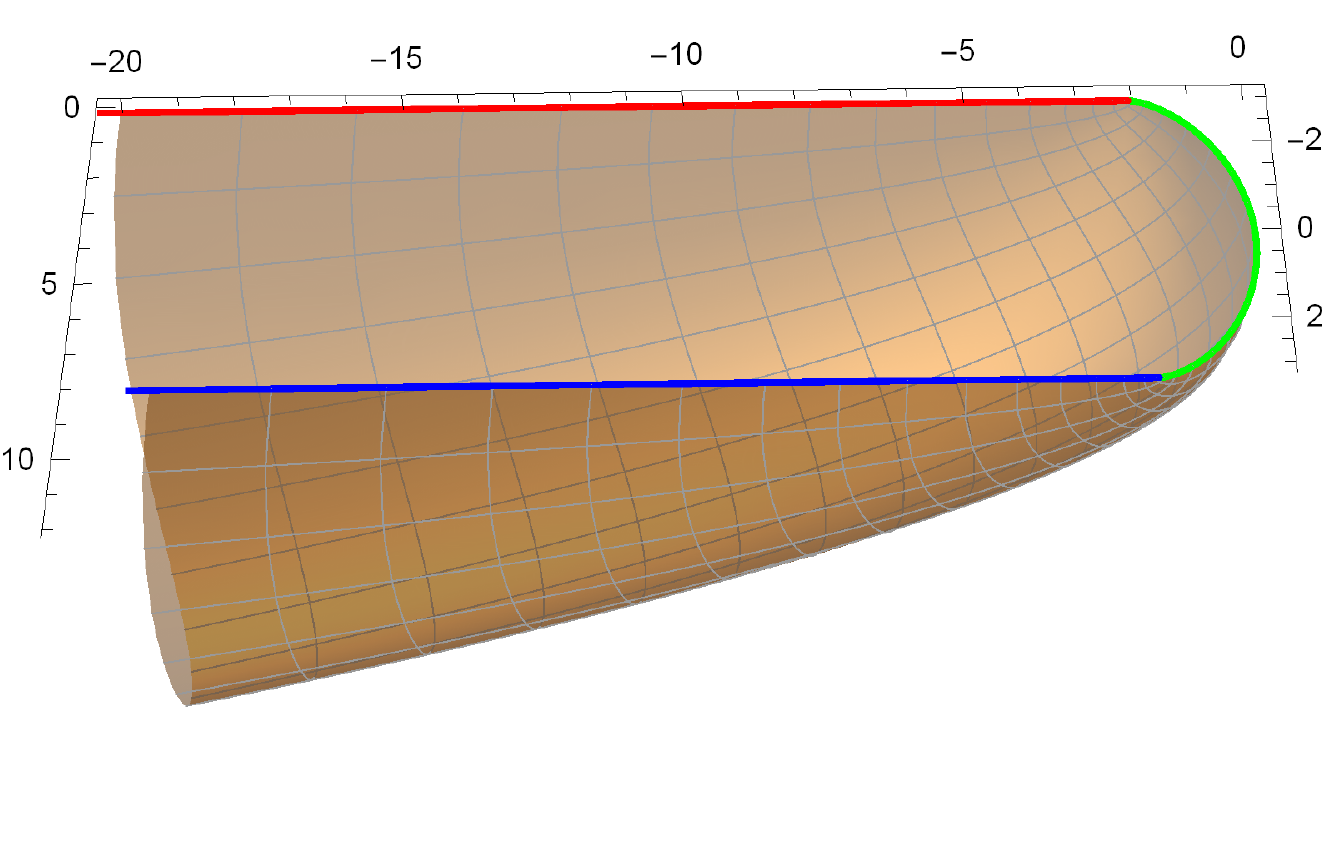}} \hspace{3mm}
    \caption{The $b = \sqrt{1-a^2},\beta = B =0$ case for (a) $a=0.25$, (b) $a=1/3$, (c) $a=0.5$, (d) $a=0.8$. The contour of the surface represents the Wilson loops contour in the $\{X_1,X_2,X_3\}$ subspace, the color of the surface represents the value of $Z$, blue is the AdS boundary and hotter colors are in the bulk. The red line corresponds to $\sigma = \frac{\pi}{2 a}$, the blue line corresponds to $\sigma = \frac{3\pi}{2 a}$ and the green line to $\tau = 0$. The spike goes to infinity and corresponds to $\tau \to \infty$. The green contour lives in the $X_1-X_3$ plane at $X_2 = 0$.
    In (e) we plot the $a \to 1$ limit.
    The red line corresponds to $\sigma = \frac{\pi}{2}$, the blue line corresponds to $\sigma = \frac{3\pi}{2}$ and the green line to $\tau = 0$ and is described by a cycloid. In contrast to (a)-(d), in (e) we plot the ${X_1,X_3,Z}$ subspace.
}
    \label{fig:b1masq}
\end{figure}

In this case the new solution simplifies to
\begin{align}
\hat{X}_1 & =~ 2 \frac{ (a \sin (\sigma ) \sin (a \sigma )+\cos (\sigma ) \cos (a \sigma ))}{\cosh\left(\sqrt{1-a^2} \tau \right)},\nonumber\\
\hat{X}_2 &=~2 a \left(\tanh \left(\sqrt{1-a^2} \tau \right)-\sqrt{1-a^2} \tau \right) ,\nonumber\\
\hat{X}_3 &=~ 2 \frac{ \sin (\sigma ) \cos (a \sigma )-a \cos (\sigma ) \sin (a \sigma )}{\cosh\left(\sqrt{1-a^2} \tau \right)},\nonumber\\
\hat{Z} &=~ 2 \sqrt{1-a^2} \cos (a \sigma ) \tanh \left(\sqrt{1-a^2} \tau \right).
\end{align}
The surface approaches the boundary when $\tau = 0$ and when $\sigma = \frac{1}{a}\left(\frac{\pi}{2}+\pi n\right)$, $n\in \mathbb{N}$,
thus it is mapped from a semi-infinite strips on the worldsheet.
The boundary contour defined by $\tau=0$, $\frac{\pi}{2a}<\sigma < \frac{3\pi}{2a}$ is mapped to the $X_1-X_3$ plane and is given by
\begin{align}
\hat{X}_1 = 2 (a \sin \sigma  \sin a \sigma +\cos \sigma  \cos a \sigma ),\quad
\hat{X}_2 = 0,\quad
\hat{X}_3 = 2 (\sin \sigma  \cos a \sigma - a \cos \sigma  \sin a \sigma) .
\end{align}
The $\tau>0$, $\sigma = \frac{\pi}{2a},\frac{3\pi}{2a}$ boundary contours are mapped to
\begin{align}
& \hat{X}_1  =~ \frac{ 2 a \sin \frac{\pi}{2 a}  }{\cosh\left(\sqrt{1-a^2} \tau \right)},\quad
\hat{X}_2 =~2 a \left(\tanh \left(\sqrt{1-a^2} \tau \right)-\sqrt{1-a^2} \tau \right) ,\quad
\hat{X}_3 =~ \frac{- 2 a \cos \frac{\pi}{2 a} }{\cosh\left(\sqrt{1-a^2} \tau \right)},\nonumber\\
& \hat{X}_1  =~ \frac{- 2 a \sin \frac{3\pi}{2 a}  }{\cosh\left(\sqrt{1-a^2} \tau \right)},\quad
\hat{X}_2 =~2 a \left(\tanh \left(\sqrt{1-a^2} \tau \right)-\sqrt{1-a^2} \tau \right) ,\quad
\hat{X}_3 =~ \frac{ 2 a \cos \frac{3\pi}{2 a} }{\cosh\left(\sqrt{1-a^2} \tau \right)}.
\end{align}
Whenever $a = 1/(2n+1)$, $n\in \mathbb{N}$ the two curves coincide,
otherwise each contour lives on a plane, where the planes intersect at an angle $\pi/a$, see figure \ref{fig:b1masq}.
These Wilson loops have three cusps, at $\tau=0,\sigma=\frac{\pi}{2a},\frac{\pi}{2a}$ and at $\tau\to \infty$.

By conformal transformation we can map the $\sigma=\frac{\pi}{2a},\frac{3\pi}{2a}$ curves to be compact and lie on the same plane, while the $\tau=0$ curve lives on an orthogonal plane.

\textbf{$a \to 1$ limit:}
We can expand around $a=1$, and keep only the terms of order $\mathcal{O}(a-1)$ which also solve the equations of motion. The result is
\begin{align}
\hat{X}_1 = -\cos 2 \sigma  +2 \tau ^2+1,\quad
\hat{X}_2 = 0,\quad
\hat{X}_3 = -\sin 2 \sigma - 2 \sigma,\quad
\hat{Z} = - 4 \tau \cos\sigma.
\end{align}
The boundary contour is given by two parallel lines connected by a cycloid, see figure \ref{fig:b1masq} (e).
The $\sigma = \pi$ profile is given by $\hat{Z}  = \sqrt{8 \hat{X}_1}$.

\subsubsection*{$a = \frac{1}{2q},\beta = 0, B =\frac{\pi(q-m)}{2q}$}

\begin{figure}
    \centering
    \subfloat[][]{
  \includegraphics[width=50mm]{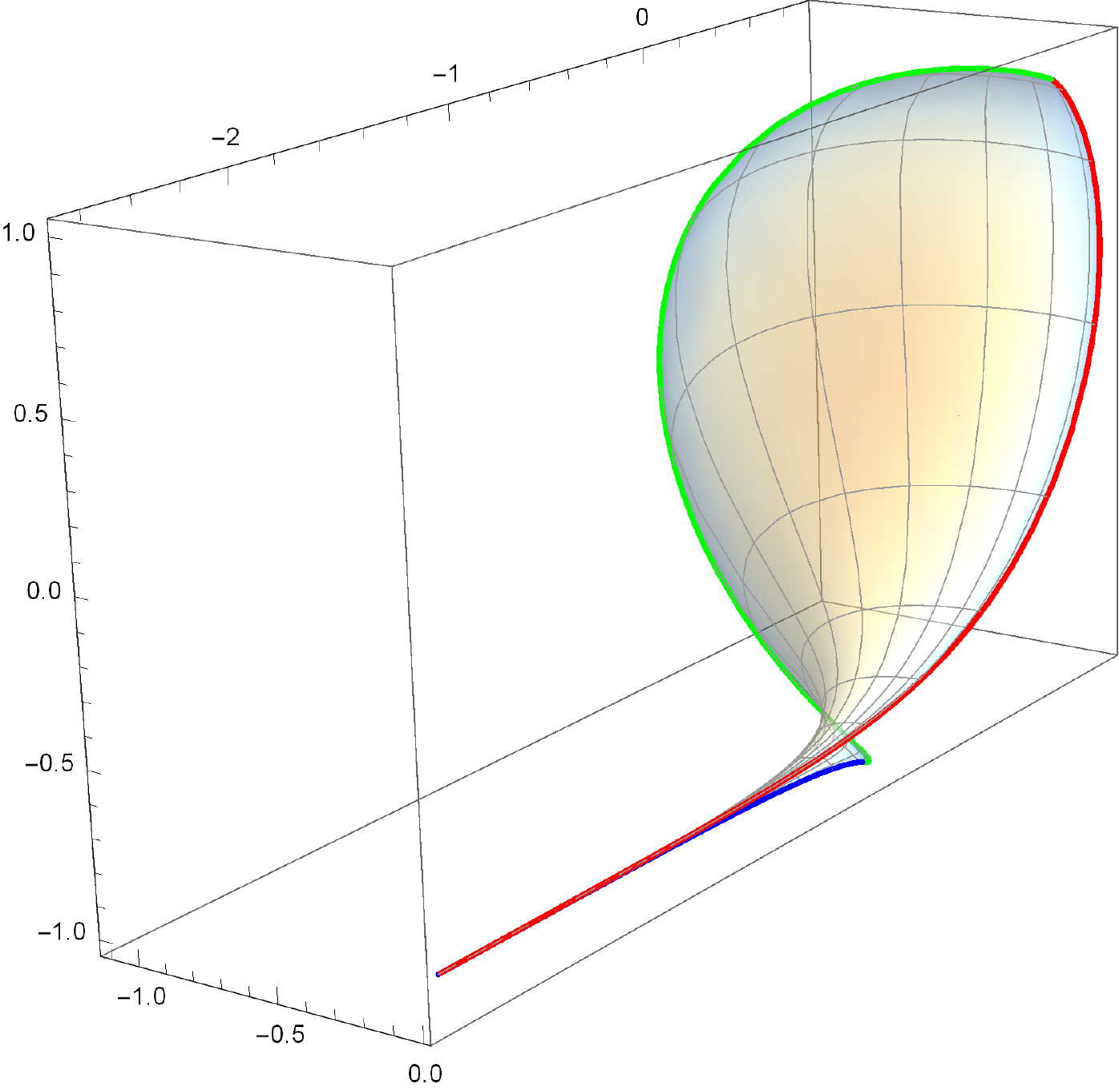}} \hspace{3mm}
    \subfloat[][]{
  \includegraphics[width=50mm]{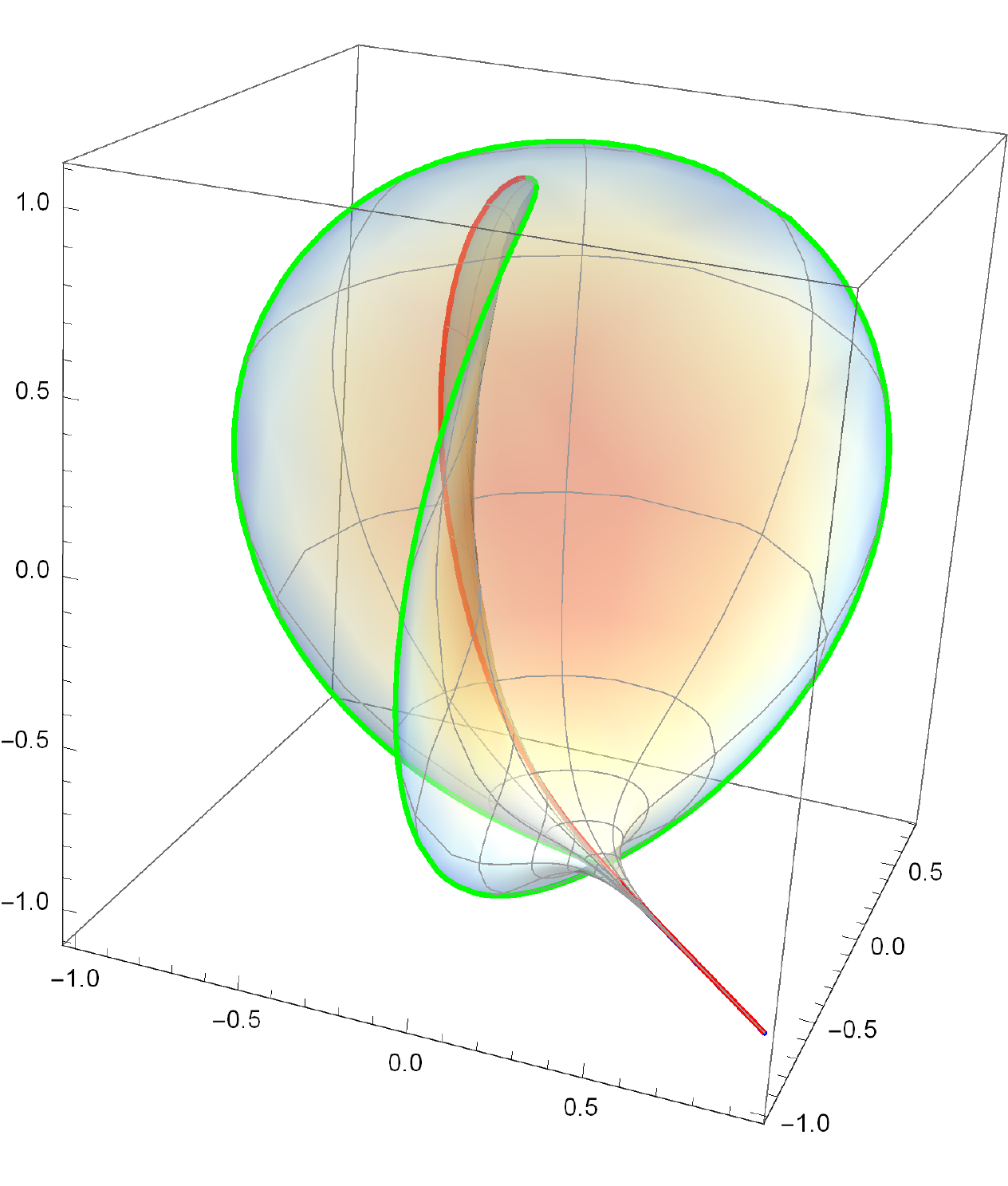}} \hspace{3mm}
    \caption{$a = \frac{1}{2 q},\beta  =0, B = B =\frac{\pi(q-m)}{2q}, b = 0.3$, for (a) $q=1,m=1$ and (b) $q=2,m=1$. The contour is the Wilson loops contour, and the red line corresponds $\sigma = \pi m$, the blue line corresponds $\sigma = \pi(m+2q)$ and the green line corresponds $\tau = \tau_{min}(\sigma)$. We plot the ${X_1,X_2,X_3}$ coordinates and the color of the surface corresponds to the value of $Z$, blue on the boundary and hotter in the bulk. As we increase $q$ the contour becomes more and more complicated.}
    \label{fig:a1o2nfullplot}
\end{figure}

In this case the boundary of the Wilson loop is made of three connected sections defined by $\sigma = m \pi$, $\sigma = m \pi +2 \pi q$ and
\begin{align}
\frac{2 \sinh ^{-1}\left(\frac{\sqrt{4-\frac{1}{q^2}} \left(4 \left(b^2-1\right) q^2+1\right) \sin (\sigma ) \csc \left(\frac{\pi  m-\sigma }{2 q}\right)}{4 b \left(4 q^2-1\right)}\right)}{\sqrt{4-\frac{1}{q^2}}}<\tau<\infty.
\end{align}
As in the previous case we have three cusps where the sections join, see figure \ref{fig:a1o2nfullplot}.

\subsubsection*{The generic case}
\begin{figure}
    \centering
    \subfloat[][]{
  \includegraphics[width=35mm]{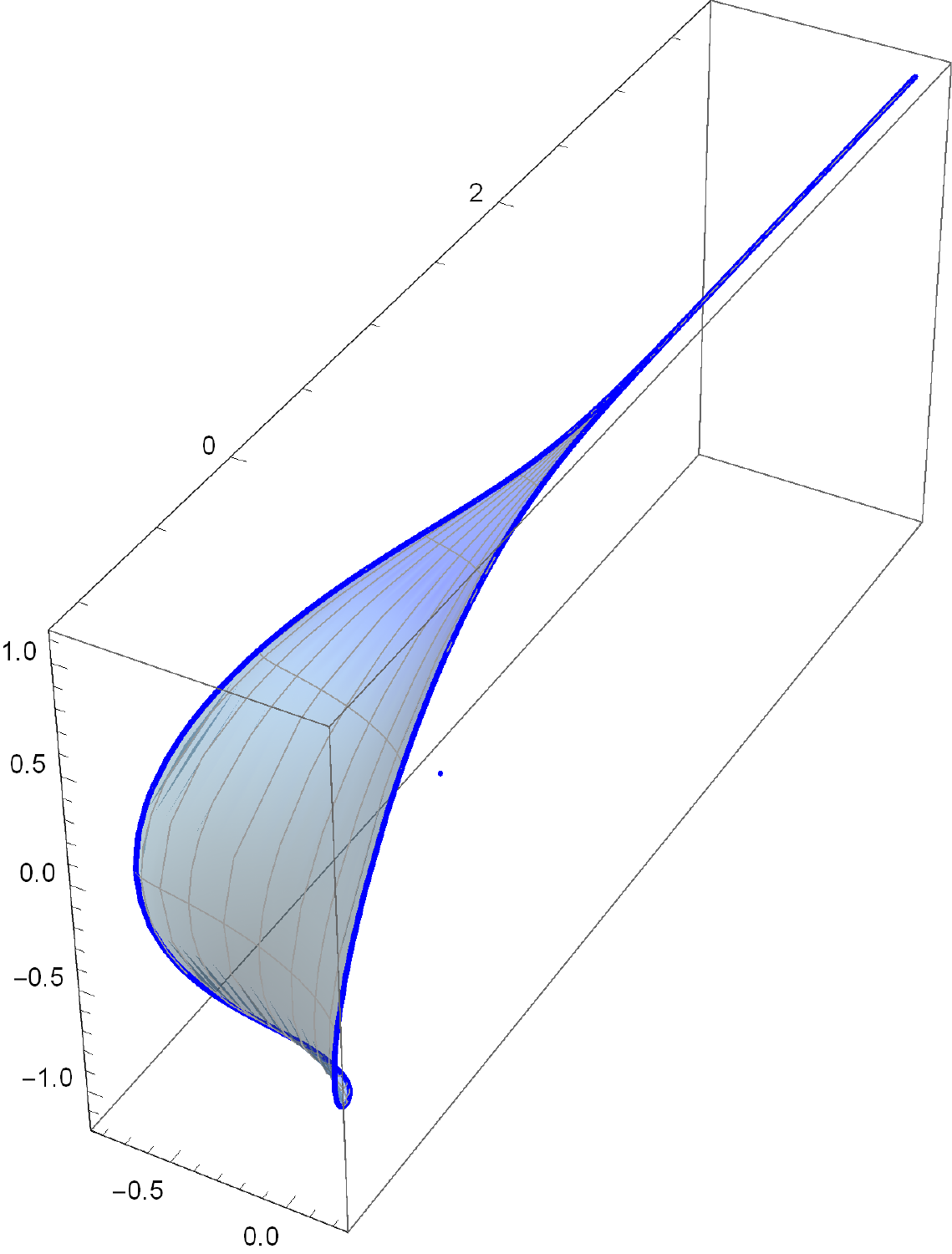}} \hspace{3mm}
    \subfloat[][]{
  \includegraphics[width=35mm]{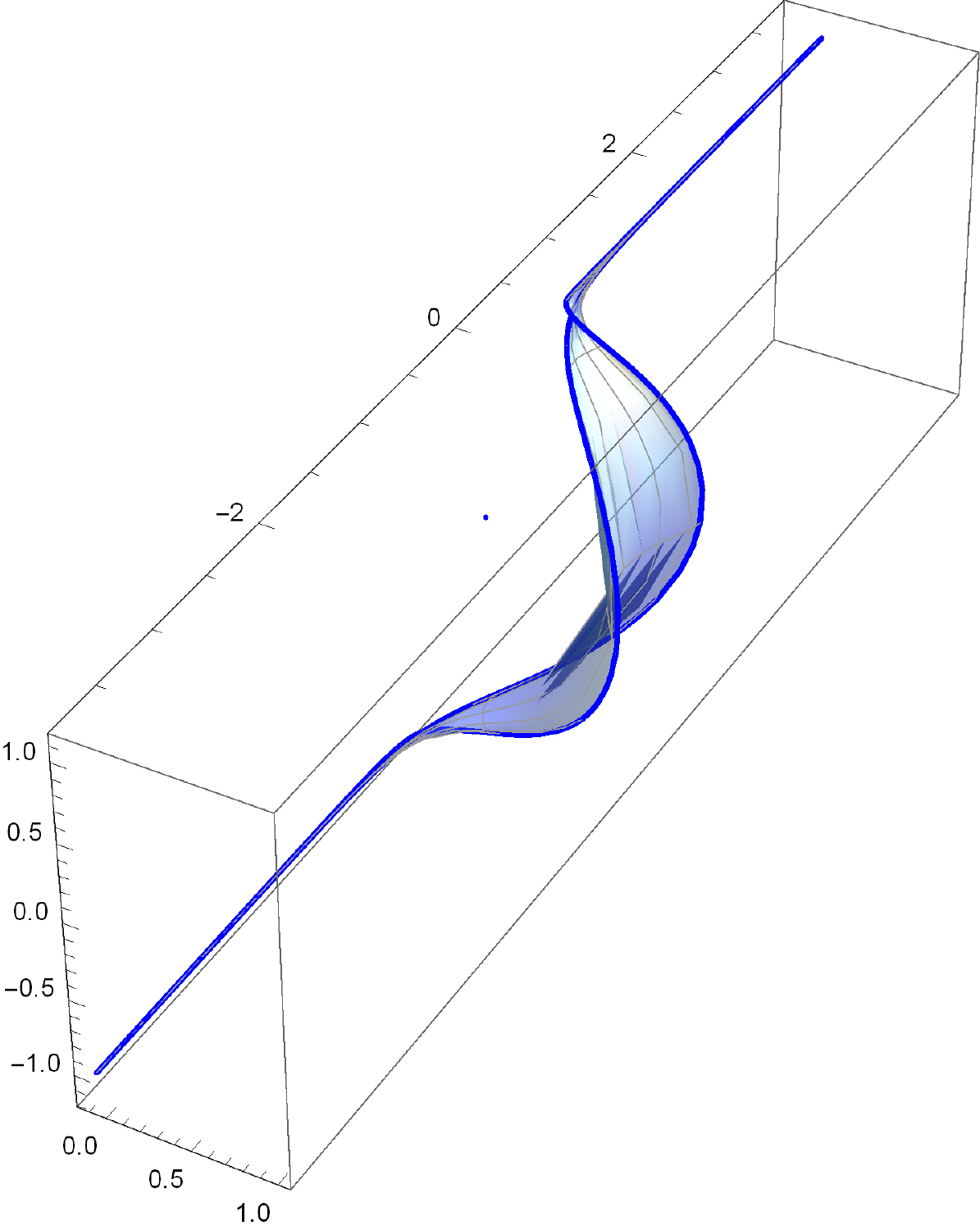}} \hspace{3mm}
    \subfloat[][]{
  \includegraphics[width=35mm]{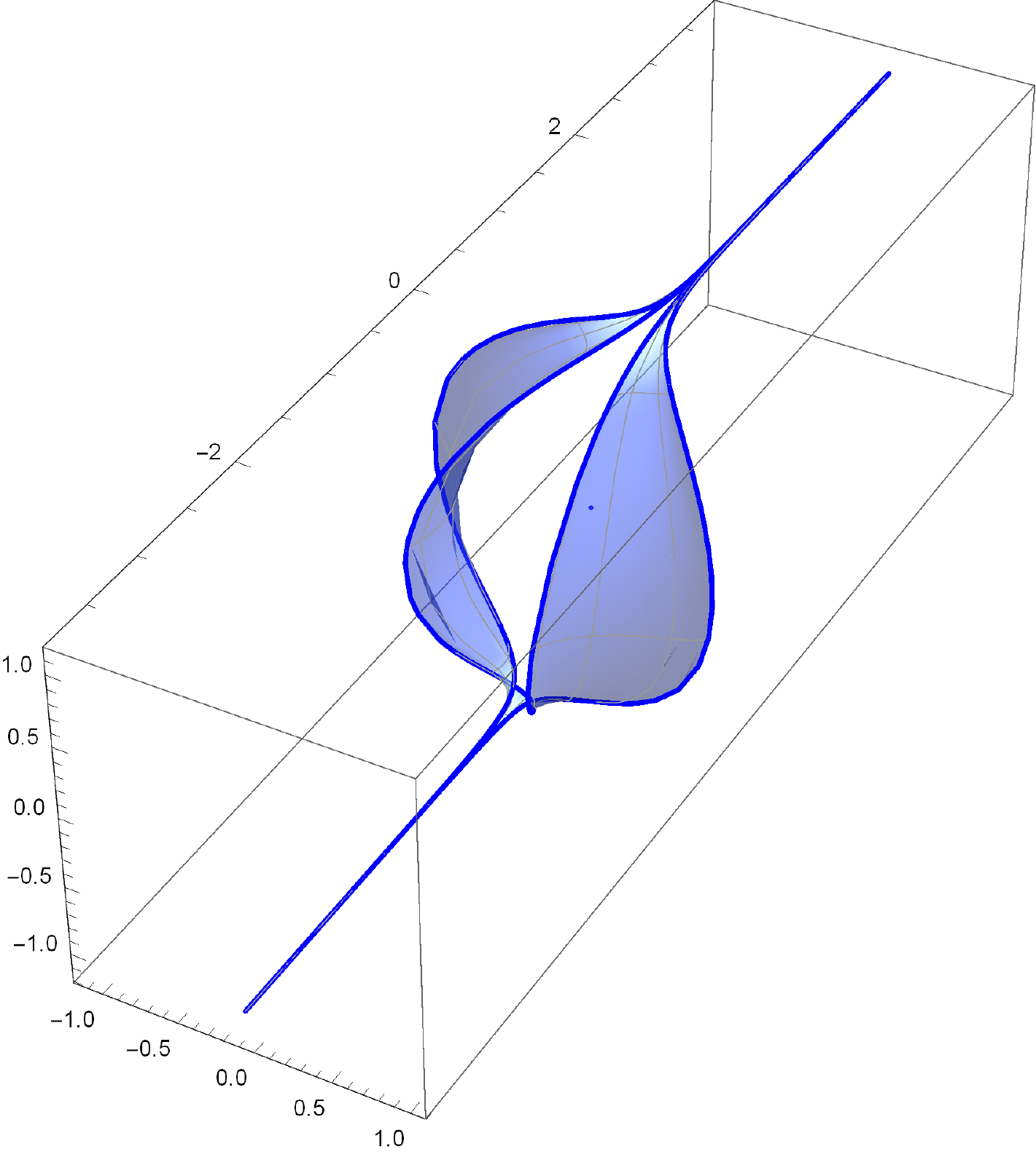}} \hspace{3mm}
    \subfloat[][]{
  \includegraphics[width=35mm]{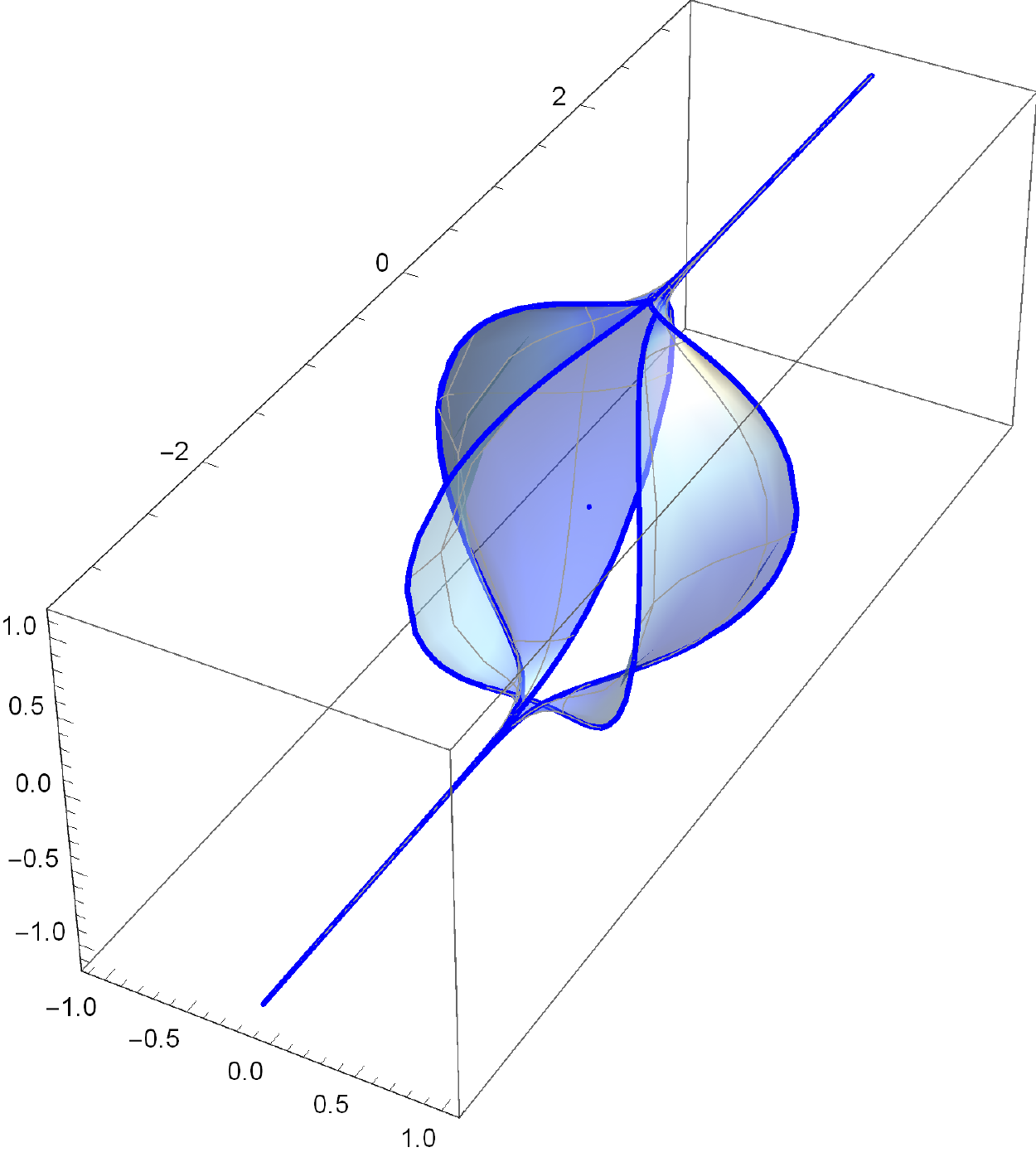}} \hspace{3mm}
    \caption{Plots of the generic case, where different number of cusps can develop. We use $a=0.7$, $b=0.3$, $\beta = 1.068$. In figures (a)-(d) we have 1-4 cusps respectively. We plot the $\{X_1,X_2,X_3\}$ space. The blue line corresponds to the boundary contour, and the color of the surface corresponds to the value of $Z$, blue on the boundary and hotter in the bulk.}
    \label{fig:longitudea07b03cusps}
\end{figure}

In the generic case we may have Wilson loops with different number of cusps $n=1,2,3,..$ depending on the parameters of the transformation, see figure \ref{fig:longitudea07b03cusps}.

\subsection{BMN geodesic}
\begin{figure}
    \centering
    \subfloat[][]{
  \includegraphics[width=50mm]{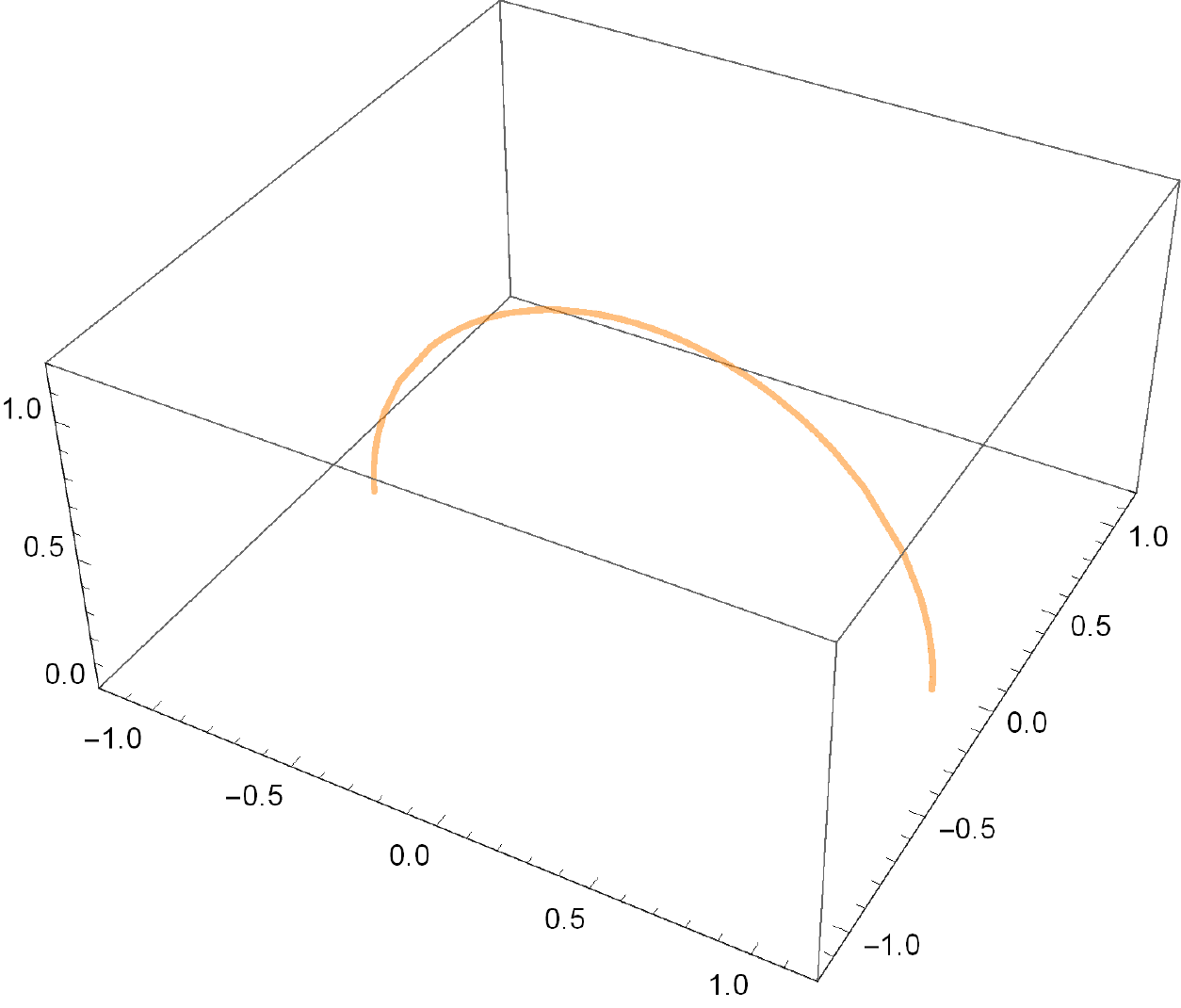}} \hspace{3mm}
    \subfloat[][]{
  \includegraphics[width=50mm]{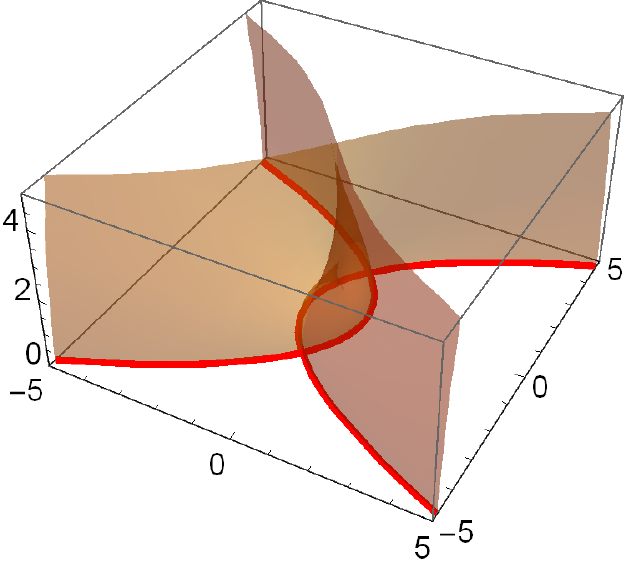}} \hspace{3mm}
    \subfloat[][]{
  \includegraphics[width=50mm]{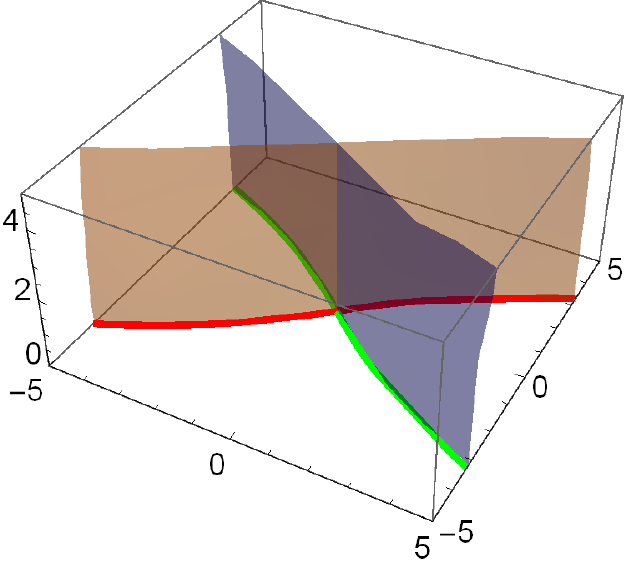}} \hspace{3mm}
    \caption{The BMN geodesic plotted in $AdS_3$. In (a) we plot the original solution in $\AdS_3$.
    After the transformation with $\beta=\pi/2$ we have (b) $b=0.3$ where we have two infinite connected contours (red), (c) $b=1.5$ where we have two infinite disconnected contours, red and green.}
    \label{fig:staticstringafterinversion}
\end{figure}

A simple string solution living in $\AdS_2\times \Sphere^1$ is given by the BMN geodesic\footnote{We could also start with a simpler configuration related to this one by a conformal transformation, $Z = e^{-\sigma}$, $X^\mu = 0$, or any other solution related by conformal transformation. The transformed solutions are not related by conformal transformations, since they do not commute with the dual conformal transformation. In any case, the simpler solution also results with a simpler surface ending on a straight line. In this section we focus on the more interesting configuration (\ref{eq:BMNsol}).}
\begin{align}\label{eq:BMNsol}
X_1 = \tanh\sigma,\quad
Z=\frac{1}{\cosh\sigma},\quad
\Phi = i \sigma,
\end{align}
with $-\infty< \sigma < \infty$, and $\phi$ is the angle of $\Sphere^1 \subset \Sphere^5$
We can easily apply the procedure to this solution and get
\begin{align}
\hat{X}_1 = &~\tanh \sigma -2 b \tau  \cos \beta  \tanh \sigma + b^2 \cos 2 \beta  \left(\tau ^2 \tanh \sigma +\sigma \right)  ,\nonumber\\
\hat{X}_2 = &~ ~~~\qquad -2 b \tau  \sin \beta  \tanh \sigma + b^2 \sin 2 \beta  \left(\tau ^2 \tanh \sigma +\sigma \right) ,\nonumber\\
\hat{Z}= &~\text{sech}\sigma -2 b \tau  \cos \beta  \text{sech}\sigma -\frac{1}{2} b^2 \text{sech}\sigma  \left(\cosh 2 \sigma -2 \tau ^2+1\right) .
\end{align}
Quite interestingly, this simple solution which approaches the boundary at two points generates a non-trivial solutions with a boundary contour.
There are two different type of solutions depending on whether $b$ is larger or smaller than $\sin\beta$ (and a marginal case when $b=\sin\beta$).
When $b < \sin\beta$ we have two infinite contours connected in the bulk, where for $b > \sin\beta$ we have two disconnected infinite boundary contours, see figure \ref{fig:staticstringafterinversion}.
The angle between the lines, which become straight at infinity, depends on $\tan\beta$.

\section{Discussion}

In this paper we applied dual conformal transformations to various holographic Wilson loops and related solutions in $\AdS_5 \times \Sphere^5$.
The way we defined the dual conformal transformations is by T-dualizing the solution along all the flat coordinates $X^\mu$ in the Poincare patch, then acting with conformal transformations in the dual space and finally T-dualizing back to the original space.
The transformation maps between solution of the equations of motion leaving the Virasoro constraints invariant, but does not leave the Lagrangian invariant.

Our initial solutions ended on the AdS boundary at $Z=0$, however T-duality does not necessarily map the boundary to the boundary\footnote{In the case of of the null cusp or four null cusps it was shown that the boundary does map to the boundary after T-duality \cite{Alday:2007hr}, however one can check that applying our procedure, T-dualizing back after the conformal transformation does not map the boundary to itself.}), so after applying a conformal transformation in the dual space and T-dualizing back, the boundary contour of the initial solution does not map back to the AdS boundary.
Consequentially we needed to analytically continue the solutions on the worldsheet, and include regions which were mapped initially to $Z<0$.
Since our solutions end on the AdS boundary, a proper regularization prescription is needed.
By regularizing the new solutions in the standard way \cite{Maldacena:1998im,Drukker:1999zq}, we found that the regularized area in general has changed.
Also, in case we keep track of the area defined with the original regulator, the expectations value would still change since the Lagrangian is not invariant.

We acted only with dual special conformal transformations since the other transformations correspond to regular conformal transformations.
In terms of the dual special conformal parameter denoted by $b^\mu$, the new solution is a second order polynomial in $b = |b^\mu|$, so the $b^2$ coefficient is also a new solution by itself.

We have seen that the transformation may result in a surface with very different features compared to the one we started with.
For example, the surface may loose or develop new cusps, and a closed boundary contour may open up and vice versa.

The problem of finding minimal surfaces in $\AdS_3$ can be Pohlmeyer reduced and the information of the solution is encoded in a holomorphic function $f(z)$ and a definition of the boundary contour in the worldsheet $z$-plane.
We find that the transformation changes the worldsheet boundary contour and leaves $f(z)$ invariant, however when comparing to the original solution we should map the worldsheet contour to the original one by a holomorphic map, which ultimately changes $f(z)$, which means the new solution will not be related to the original one by a conformal transformation.

Dual conformal transformations are usually associated with the Yangian symmetry, and are expected to leave the expectation value invariant.
The Yangian symmetry was also shown recently to be related to the spectral parameter deformation of smooth Wilson loops in \cite{Klose:2016qfv,Klose:2016uur}.
However, the transformations introduced in this paper do change the expectation value,
and do not generate the spectral parameter deformation as we checked explicitly.
It would be very interesting to find out whether a different regularization scheme, or some modification of the procedure can generate the these symmetries.

Our procedure can be viewed as a solution generating technique.
There are other well known integrability methods of generating new solution form a given solution, such as the dressing method \cite{Zakharov:1973pp} for example.
The dressing method was applied to the longitude solution in \cite{Kalousios:2011hc}, which we have also studied here.
Our generated solutions do not seem to be related to the ones generated in \cite{Kalousios:2011hc} in an obvious way.
It would be interesting to clarify whether this is indeed the case, or if there is any relation between the methods, or other known methods.

In this paper we used the self-duality property of the AdS background under T-duality along the flat AdS directions.
There are other sequences of T-duality which leave the $\AdS_5 \times \Sphere^5$ background invariant which involve a formal T-duality along coordinates of the sphere \cite{Berkovits:2008ic,Dekel:2011qw}. One can try to apply our procedure using using these T-dualities combined with dual symmetry transformations of the sphere to generate more solutions from well known solutions such as the latitude \cite{Drukker:2007qr} (see also \cite{Forini:2015bgo,Faraggi:2016ekd} for recent analysis) or a correlation function of the latitude with the BMN operator \cite{Giombi:2012ep}.

It would also be interesting to check how the algebraic curve associated to the solutions changes under the transformation.
In general it is hard to compute the algebraic curve, however for our initial solutions it could be easily done using the method of \cite{Dekel:2013dy}, by computing the Lax operator directly.
Once we have the explicit Lax operator for the initial solution, one might be able to compute the Lax operator to leading order in $b$, by expanding the Lax equations.
Another possibility for the $\AdS_3$ solutions would be to try to use the general known solutions in terms of Riemann theta-functions \cite{Babich:1992mc,Ishizeki:2011bf} where the Lax operators was constructed explicitly in \cite{Cooke:2014uga}, apply the transformation and try to identify the new solution in terms of the general solution.

Throughout the paper we applied the dual conformal transformation once on an initial solution.
It would be interesting to understand the action of a sequence of such transformations.
Denoting the procedure schematically as $(TcT)$, acting again with a dual conformal transformation will give
$(TcT)\times (Tc'T) = (T cc'T) = (T\tilde cT)$, namely the same transformation with different parameters.
A more interesting thing would be to act with a sequence of the form $(cTc'T)\times(c''Tc'''T)...$, and find out whether this group of transformations has some interesting algebraic structure, and if it generates a finite or infinite parameter family of solutions.

\section*{Acknowledgments}

I would like to thank T. Bargheer, N. Beisert, N. Drukker, M. Heinze, T. Klose, G. Korchemsky, M. Kruczenski, F. Loebbert, T. McLoughlin, H. Munkler, A. Sever, C. Vergu, E. Vescovi, K. Zarembo, and the participants and organizers of the Focus program in Humboldt U. Berlin 2016, where some of the results were presented, for valuable discussion and comments.
I also thank N. Drukker and K. Zarembo for comments on the manuscript.
This work was supported by the Swedish Research Council (VR) grant 2013-4329.

\appendix

\section{Longitude transformation}\label{ap:long}
In this appendix we give the transformed longitude solution,
\scriptsize
\begin{align}
x_1 &=
\frac{a \sin  \sigma   \sin  a \sigma  +\cos  \sigma   \cos  a \sigma  }{\cosh\tilde a \tau}
+\frac{2 b \left(\sin \beta \text{sech}\tau  \tilde{a} (a \cos \sigma  \sin a \sigma -\sin \sigma \cos a \sigma )+a \cos \beta \sin B\left(\tau  \tilde{a}-\tanh \tau  \tilde{a}\right)\right)}{\tilde{a}}\nonumber\\
&-\frac{b^2 }{\tilde{a}^2\cosh\tau  \tilde{a} }\bigg[
-\sin 2 \beta \cos B\sinh \tau  \tilde{a}
-a \sin ^2\beta \sin \sigma \sin a \sigma
-\sin ^2\beta \cos \sigma \cos a \sigma
+\cos ^2\beta \cos 2 B\cos \sigma \cos a \sigma \nonumber\\
&-\cos ^2\beta \sin 2 B\cos \sigma \sin a \sigma
+a \cos ^2\beta \sin \sigma \sin (a \sigma +2 B)
\bigg],\nonumber\\
x_2 &= \frac{a \sin \sigma \cos a \sigma -\cos \sigma \sin a \sigma }{\cosh\tau  \tilde{a}}
+\frac{2 b \left(\sin \beta \text{sech}\tau  \tilde{a}(\sin \sigma \sin a \sigma +a \cos \sigma \cos a \sigma )+a \cos \beta \cos B
\left(\tanh \tau  \tilde{a}-\tau  \tilde{a}\right)\right)}{\tilde{a}},\nonumber\\
&+\frac{b^2 }{\tilde{a}^2\cosh\tau  \tilde{a} }\bigg[
\sin ^2\beta (\cos \sigma \sin a \sigma
-a \sin \sigma \cos a \sigma )+\cos ^2\beta (\cos \sigma
\sin (a \sigma +2 B)-a \sin \sigma \cos (a \sigma +2 B))\nonumber\\
&-\sin 2 \beta \sin B\sinh \tau  \tilde{a}
\bigg],\nonumber\\
x_3 &= -\tanh \tau  \tilde{a}
+\frac{2 b \cos \beta (\sin \sigma \cos (a \sigma +B)-a \cos \sigma \sin (a \sigma +B))}{\tilde{a}\cosh\tau  \tilde{a}}\nonumber\\
&+\frac{b^2 }{2\tilde{a}^2\cosh\tau  \tilde{a} }\left(2 \cos 2 \beta \sinh \tau  \tilde{a}+\sin 2 \beta ((a+1) \cos ((a-1) \sigma +B)-(a-1) \cos (a \sigma +B+\sigma ))\right),\nonumber\\
z &  = \frac{\tilde{a} \sin \sigma  }{\cosh\tau  \tilde{a}}
+\frac{2 b \left(\cos \beta \sinh \tau  \tilde{a}\cos (a \sigma +B)+\sin \beta \cos \sigma \right)}{\cosh\left(\tau  \tilde{a}\right) }
-\frac{b^2 \sin \sigma }{\tilde{a}\cosh\tau  \tilde{a}},
\end{align}
where we defined $\tilde a \equiv \sqrt{1-a^2}$.
The $b\to \infty$ limit yields again a longitude solution.

\bibliographystyle{nb}
\bibliography{refs}

\end{document}